\documentclass{article}

\usepackage{arxiv}

\usepackage[utf8]{inputenc} 
\usepackage[T1]{fontenc}    
\usepackage{hyperref}       
\usepackage{url}            
\usepackage{booktabs}       
\usepackage{amsfonts}       
\usepackage{nicefrac}       
\usepackage{microtype}      
\usepackage{lipsum}
\usepackage{graphicx}
\graphicspath{ {./images/} }

\usepackage{longtable}
\usepackage{amsmath}
\usepackage{bm}
\usepackage{soul}
\newcommand{\mi}{{\rm i}}
\newcommand{\md}{{\rm d}}
\newcommand{\alpA}{\alpha_{\rm A}}
\newcommand{\alpB}{\alpha_{\rm B}}
\newcommand{\alpC}{\alpha_{\rm C}}
\newcommand{\ra}{{\bm{r}}_{\rm A}}
\newcommand{\rb}{{\bm{r}}_{\rm B}}

\title{Effective screening of medium-assisted Van der Waals interactions between embedded particles}

\author{
 Johannes Fiedler\thanks{Department of Physics and Technology, University of Bergen, All\'egaten 55, 5020 Bergen, Norway.} \\
  Institute of Physics\\University of Freiburg\\Hermann-Herder-Str. 3, 79104 Freiburg, Germany \\
  \texttt{johannes.fiedler@uib.no} \\
   \And
 Michael Walter\thanks{Cluster of Excellence livMatS @ FIT} \thanks{Frauenhofer IWM, MikroTribologie Centrum $\mu$TC, W\"ohlerstrasse 11, 79108 Freiburg, Germany} \\
  FIT Freiburg Centre for Interactive\\ Materials and Bioinspired Technologies, \\Georges-K\"ohler-Allee 105, 79110 Freiburg, Germany \\
  \And
 Stefan Yoshi Buhmann\thanks{Institut f\"ur Physik, Universit\"at Kassel, Heinrich-Plett-Str. 40, 34132 Kassel, Germany} \\
  Institute of Physics\\University of Freiburg\\Hermann-Herder-Str. 3, 79104 Freiburg, Germany
}

\begin{document}
\maketitle
\begin{abstract}
The effect of an implicit medium on dispersive interactions of particle pairs is discussed and simple expressions for the correction relative to vacuum are derived. We show that a single point Gauss quadrature leads to the intuitive result that the vacuum van der Waals $C_6$ coefficient is screened by the permittivity squared of the environment evaluated near to the resonance frequencies of the interacting particles. This approximation should be particularly relevant if the medium is transparent at these frequencies. In the manuscript, we provide simple models and sets of parameters for commonly used solvents, atoms and small molecules.
\end{abstract}


\section{Introduction}

Van der Waals forces are the fundamental interactions between two neutral and polarisable particles~\cite{London1937,VanderWaals1873,London1930}.
These forces prevail in holding together many materials and play an important role in living organisms, such as geckos walking on smooth surfaces~\cite{Autumn2000}.
They have also found increasing importance in technological applications such as microelectromechanical and nanoelectromechanical components~\cite{DelRio2005}. During recent years these forces have been well-studied in several experiments~\cite{Grisenti1999,Arndt1999,Juffmann2012,Brand2015} and in theory~\cite{Casimir48,pitaevskii,Scheel2008,Buhmann12a}.

Despite their Coulombic origin dispersive forces are among the weakest forces in nature. Time-dependent perturbation theory suggests their 
interpretation as being caused by ground-state fluctuations of the electromagnetic fields. This view has been taken in  Casimir theory~\cite{Casimir48} dealing with two dielectric plates in vacuum as well as in colloidal systems~\cite{McLachlan387}, namely the stabilisation of hydrophobic suspensions of particles in dilute electrolytes~\cite{MILONNI1994217}.
Alternative approaches derive dispersion forces from position-dependent ground-state energies of the coupled field--matter system~\cite{London1937,Casimir48,Casimir482}.
These descriptions are restricted to partners interacting in vacuum.
Natural systems such as colloids or proteins are often embedded in and environment such as a solvent or a matrix.

\begin{figure}[htbp]
    \centering
    \includegraphics[width=0.6\columnwidth]{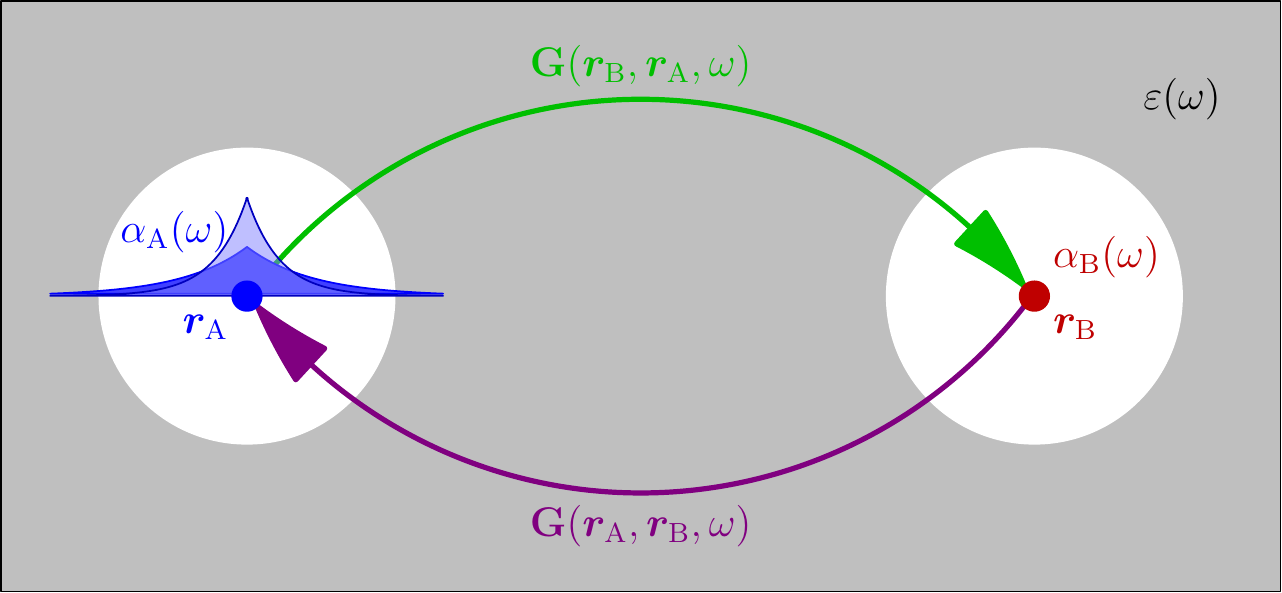}
    \caption{Illustration of two particles (with polarizabilities $\alpha_{\rm A,B}(\omega)$ located 
    at positions $\bm{r}_{\rm A,B}$. The particles are embedded in a medium with permittivity $\varepsilon(\omega)$ (grey area) where
    a medium excluded area surrounding each particle is formed.
    Interactions described by the Greens functions $\bf{G}$ are indicated. See text for details.
    }
    \label{fig:schematicfigure}
\end{figure}

The impact of an effective medium on dispersive interactions of two particles ${\rm A,B}$ is illustrated in Fig.~\ref{fig:schematicfigure}. We adopt the simplification of considering point particles characterised by their frequency-dependent polarizabilities $\alpha_{\rm A,B}(\omega)$ embedded in an effective medium characterised by its frequency-dependent permittivity $\varepsilon(\omega)$. In this picture the Van der Waals potential may be expressed as\cite{Buhmann12a}
\begin{align}
   U({\bm{r}}_{\rm A},{\bm{r}}_{\rm B}) = -\frac{\hbar \mu_0^2}{2\pi}\int\limits_0^\infty \mathrm d \xi \, \xi^4 \operatorname{Tr}\left[ \alpha_{\rm A}(\mi\xi)\cdot{\bf{G}}({\bm{r}}_{\rm A},{\bm{r}}_{\rm B},\mi\xi)\cdot\alpha_{\rm B}(\mi\xi)\cdot {\bf{G}}({\bm{r}}_{\rm B},{\bm{r}}_{\rm A},\mi\xi)\right]\,,\label{eq:UVDW}
\end{align}
where the Green functions $\bf{G}$ represent the properties of the field including the medium. One may picturize $\bf{G}$ as describing the interaction between both particles via the exchange of virtual photons. Equation~(\ref{eq:UVDW}) has to be read from right to left: a virtual photon $\mi\xi$ is created at position $\ra$ and propagates to particle B, which is expressed by the Green function ${\bf{G}}(\rb,\ra,\mi\xi)$. At this point it interacts with the polarizability of particle B, $\alpB(\mi\xi)$ and is back-scattered to particle A, again expressed by the Green function ${\bf{G}}(\ra,\rb,\mi\xi)$, where it interacts with particle A. The sum (integral) over all possible virtual photons yields the total Van der Waals interaction.

The presence of a medium as the environment has two distinct effects influencing dispersive interactions between A and B:
\begin{itemize}
    \item[(I)] \textit{Deformation of the particle's electron density:} 
    Caused by the short distances between the considered particles and the environmental particles, its wave function is modified compared to the one of the free particle~\cite{doi:10.1063/1.1901584}. This phenomenon is depicted for particle A in Fig.~\ref{fig:schematicfigure} by the blue (probability of presence) for the free particle and the semitransparent blue area for the confined particle. This deforming effect affects the polarizabilities of both particles $\alpha_{\rm A,B}(\mi\xi)$ in the Van der Waals interaction~(\ref{eq:UVDW}).
    \item[(II)]\textit{Screening of the virtual photon's propagation:} In Figure~\ref{fig:schematicfigure}, it can be observed that the virtual photon has to pass the medium. This can be approximated~\cite{Johannes} to lead to a damping by $1/\varepsilon(\mi\xi)$ for each propagation direction. This process leads to the excess polarizability models~\cite{Johannes,Sambale2007}.
\end{itemize}
Applying the above to Eq.~(\ref{eq:UVDW}) for a bulk material, the medium-assisted Van der Waals interaction between two particles A and B embedded in a medium with permittivity $\varepsilon(\omega)$ separated by the distance $d$ in the nonretarded limit reads~\cite{Johannes,Buhmann12a,McLachlan387}
\begin{equation}
    U_{\rm vdW}(d) = -\frac{C_6}{d^6} \, ,\quad C_6 = \frac{3\hbar}{16\pi^3 \varepsilon_0^2}\int\limits_0^\infty \frac{\alpA^\star(\mi \xi)\alpB^\star(\mi \xi) } {\varepsilon^2(\mi \xi)}\md \xi \, , \label{eq:C6}
\end{equation}
with the reduced Planck constant $\hbar$ and the vacuum permittivity $\varepsilon_0$. The $\alpha^\star_{\rm A,B}(\mi\xi)$ are understood to be modified by the presence of the medium.

The procedure to estimate medium-assisted dispersion interactions by the integral~(\ref{eq:C6}) is challenging in practical calculations as the polarizabilities as well as the permittivity have to be known over the full frequency range. Furthermore, this integral can get very complex depending on the environmental medium. For instance, the most-commonly applied medium is water, whose currently most exact model consists of 19 damped oscillators, 7 for the infrared and 12 for the ultraviolet regime and 2 Debye terms are involved in order to match the experimental data in the low-frequency regime~\cite{Fiedler2020}.

Practical electronic structure calculations in the spirit of the model presented in Fig.~\ref{fig:schematicfigure} describe the environment by a polarizable continuum model (PCM)~\cite{Tomasi2005,Held2014} based on the static permittivity sufficient for ground state calculations. The workhorse of electronic structure theory is density functional theory (DFT). The most common functional approximations within DFT are known to severely lack the description of dispersion interactions. This can be corrected by modifying the functional~\cite{Berland2015} or by adding a dispersive correction to the energy. The latter is in the spirit of our considerations applied by several approaches like the semi-empirical Grimme~\cite{Grimme2010,Caldeweyher2017} or the Tkatchenko--Scheffler~\cite{Tkatchenko2009,Tkatchenko2012} models. 
Such models are directly applicable to describe the presence of an explicit environment where all solvent molecules are resolved.
Newer developments even take many body interactions into account\cite{Tkatchenko2012,Caldeweyher2017} and should therefore be capable of including nontrivial environmental screening effects  at least partly.\cite{https://doi.org/10.1002/qua.24635}
An explicit description of the environment is computationally very demanding and requires averaging over many different configurations of the environment, e.g. by explicit time propagation.\cite{Dopieralski13}

Form the viewpoint of an implicit approach like the PCM, these corrections are based on free-particle interactions disregarding the presence of the environment. While appropriate in case that both interacting particles are within the same cavity~\cite{Takahashi2019} this approach disregards screening by an implicit environment. It was therefore suggested to scale the van der Waals contributions by $\varepsilon^{-2}(\omega)$ with $\omega$ in the optical range~\cite{Hartl2020}.

We rationalize this conjecture by presenting an algebraic approximation for the medium-assisted $C_6$ coefficient 
in the following. It is based on a one-point Gauss quadrature rule leading to
\begin{equation}
    C_6^{\rm app} = \frac{C_6^{\rm AB}}{\varepsilon^2(\mi\overline{\omega})}=\left(\frac{3\varepsilon(\mi\overline{\omega})}{1+2\varepsilon(\mi\overline{\omega})}\right)^4\frac{C_6^{\rm vac}}{\varepsilon^2(\mi\overline{\omega})} \,,\label{eq:eff:omega}
\end{equation}
with an averaged main-frequency $\overline{\omega}$ to be determined in what follows. We furthermore show that $\varepsilon^{-2}(\mi\overline{\omega})$ might be replaced by $\varepsilon^{-2}(\overline{\omega})$ in the absence of resonances of the environment at the frequency $\overline{\omega}$. The prefactor on the right hand side of Eq.~(\ref{eq:eff:omega}) denotes the transition through the interface between the vacuum bubble and the environmental medium according to the Onsager model~\cite{doi:10.1021/ja01299a050} which is the most-simplest and commonly used excess polarizability model~\cite{Johannes}
\begin{equation}
\alpha^\star(\mi\xi) = \left(\frac{3\varepsilon(\mi\xi)}{1+2\varepsilon(\mi\xi)}\right)^2\alpha(\mi\xi)\,.    
\end{equation}
Due to the resulting linearity between the free-space and the approximated van der Waals coefficient, the impact of excess polarizabilities can be easily included by means of Eq.~(\ref{eq:eff:omega}) and will not be considered explicitly in the following.

\begin{figure}[t]
    \centering
    \includegraphics[width=0.6\columnwidth]{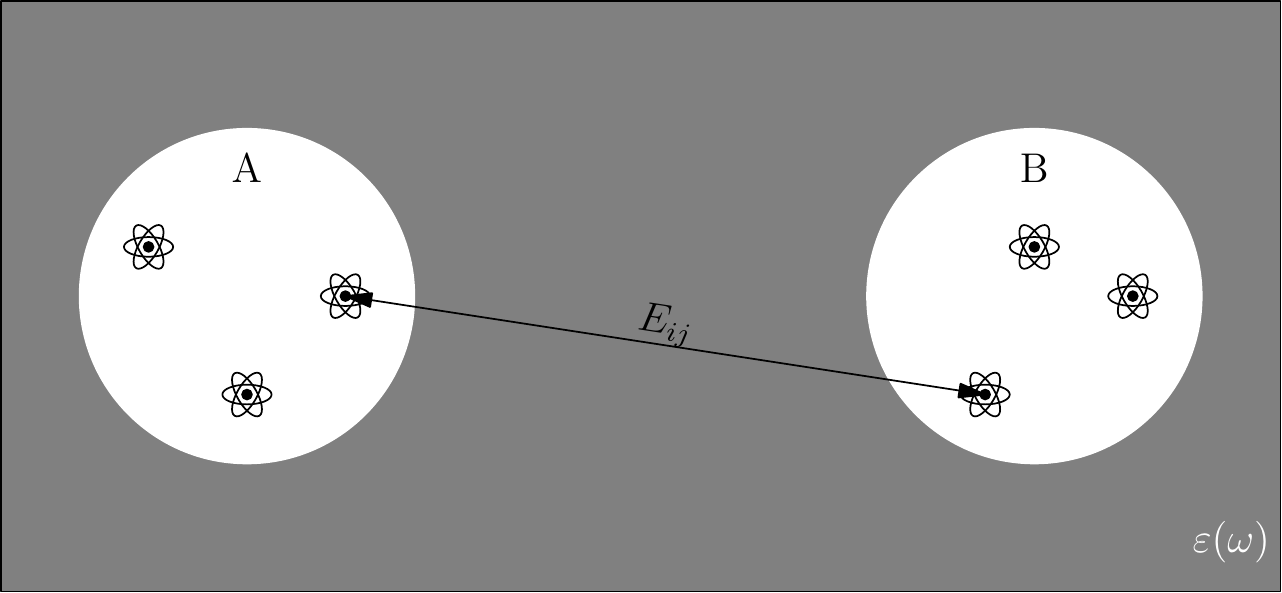}
    \caption{Schematic illustration of the microscopic picture: two ensembles of atoms A and B are embedded in a dielectric medium with permittivity $\varepsilon(\omega)$. The dispersion interaction between the ensembles via the two alternative descriptions as compact objects (mesoscopic picture, see Fig.~\ref{fig:schematicfigure}) or via the pairwise interaction as depicted here $E_{ij}$ is screened by the solvent medium.}
    \label{fig:micro}
\end{figure}
In this manuscript, we adapt the mesoscopic model to the microscopic models applied, for instance, in DFT simulations. The envisioned medium-assisted situation is depicted in Fig.~\ref{fig:micro}. Two particle ensembles A and B are embedded within a medium with permittivity $\varepsilon(\omega)$ screening the interaction. In models based on electronic structure theory, this interaction is written in the generalised Casimir--Polder form~\cite{https://doi.org/10.1002/qua.24635}
\begin{align}
    E_{\rm AB} = -\frac{\hbar}{32\pi^3\varepsilon_0^2}\int\limits_0^\infty \mathrm d \xi\int\mathrm d {\bm{r}}_{\rm A}\mathrm d {\bm{r}}_{\rm A}'\mathrm d {\bm{r}}_{\rm B}\mathrm d {\bm{r}}_{\rm B}'\frac{e^2}{\varepsilon(\mi\xi)\left|{\bm{r}}_{\rm A}-{\bm{r}}_{\rm B}\right|}\frac{e^2}{\varepsilon(\mi\xi)\left|{\bm{r}}_{\rm A}'-{\bm{r}}_{\rm B}'\right|}\chi^{\rm A}({\bm{r}}_{\rm A},{\bm{r}}_{\rm A}',\mi\xi)\chi^{\rm B}({\bm{r}}_{\rm B},{\bm{r}}_{\rm B}',\mi\xi)\,,\label{eq:GCP}
\end{align}
which describes the total dispersion energy between the systems A and B, expressed by the electronic density-density responses $\chi^i$ for $i={\rm A,B}$. In the presence of a separating medium, these are coupled via the screened Coulomb interaction $e^2/\left(4\pi\varepsilon_0\varepsilon \left|{\bm{r}}_{\rm A}-{\bm{r}}_{\rm B}\right|\right)$.
Applying the dipole approximation to (\ref{eq:GCP}), the generalised Casimir--Polder energy gets equivalent to the mesoscopic model obtained via macroscopic quantum electrodynamics Eq.~(\ref{eq:C6})
\begin{align}
E_{\rm AB} &= -\frac{C_6^{\rm AB}}{R^6} \,,\\
C_6^{\rm AB}&= \frac{3\hbar}{16\pi^3\varepsilon_0^2} \int\limits_0^\infty \mathrm d \xi \frac{\alpha^{\rm A}(\mi\xi)\alpha^{\rm B}(\mi\xi)}{\varepsilon^2(\mi\xi)}\,,\label{eq:C6DFT}\\
\alpha^{i}(\mi\xi) &= \int \mathrm d {\bm{r}}\mathrm d{\bm{r}}' \, {\bm{r}}{\bm{r}}'\chi^i({\bm{r}},{\bm{r}}',\mi\xi)\label{eq:alpha}\,,
\end{align}
where $\alpha^{i}(\mi\xi)$ denotes the screened polarizability caused by the deformation of the particle's electron density $\chi^i$ due to the presence of the environment, which does not include the effect described by the excess polarizabilities.

Typically, this interaction is separated into the pairwise interaction of the constituents (the atoms) of systems A and B, as depicted in Fig.~\ref{fig:micro}
\begin{equation}
    E_{\rm AB}=\sum_{i,j} E_{i,j}=-\sum_{i,j} f_{ij}(R_{ij})\frac{C_6^{ij}}{R_{ij}^6}\,,
\end{equation}
with $R_{ij}$ denoting the distance between atom $i$ of cloud A and atom $j$ of cloud B, and a correction function $f_{ij}(R_{ij})$ to take short-range phenomena into account. This pairwise separation of the dispersion energy corresponds to the Hamaker approach~\cite{Parsegian} (or first-order Born series expansion~\cite{Buhmann12b}) in macroscopic quantum electrodynamics. 
Such models are commonly used in modern van-der-Waals--density-functional-theory simulations with tabled vacuum $C_6$-coefficients for the different interacting constituents. 

Interestingly, within this approach the deformation of the particle's electron density (I), is expressed via a reduction of the particle's volume, which, due to the transitivity of the polarizability~(\ref{eq:alpha}), can directly be expressed by a $C_6^{\rm def}$-coefficient for the deformed electron density\cite{Tkatchenko2009}
\begin{equation}
    C_6^{\rm def} = \left(\frac{V^{\rm def}}{V^{\rm free}}\right)^2 C_6^{\rm free}\,, \label{eq:deformation}
\end{equation}
with the reduced particle volume $V^{\rm def}$ and the particle volume and $C_6$-coefficient of the free particle, $V^{\rm free}$ and $C_6^{\rm free}$, respectively. 
This assumption is questionable from the macroscopic point of view, as, for instance, the mixing of particle states near interfaces~\cite{Ribeiro2015} cannot be expressed in such simple way.
Further developments\cite{Tkatchenko2012,Caldeweyher2017} take into account similar problems due to non-additivities of the van der Waals interaction~\cite{https://doi.org/10.1002/qua.24635}. 

It can be observed that the $C_6$-coefficient~(\ref{eq:C6DFT}) depends on dispersion of the implicit environmental medium via an integration along the imaginary frequency axis. This fact motivated us to develop a simple model that takes into account the screening of the van der Waals interaction with a similar numerical effort as ordinary DFT simulations in vacuum would require. As the deformation of the particle's electron density in commonly considered in the form of Eq.~(\ref{eq:deformation}), the local-field corrections as expressed via excess polarizability models~\cite{Johannes} in the form of Eq.~(\ref{eq:eff:omega}), we neglect the explicit consideration of these effects within this manuscript.

\section{Approximation of medium-assisted C$_6$-coefficients by Gaussian quadrature}\label{sec:two}

The integral over the imaginary frequency axis for the C$_6$-coefficient~(\ref{eq:C6}) can be carried out by using a single-point Gauss quadrature rule~\cite{Parsegian,MacDowell2019}.
This method approximates the integral $I$ by 
\begin{equation}
    I=\int\limits_0^\infty f(x) g(x) \mathrm d x 
    \approx f(x_0) m_0 \,.
    \label{eq:GaussQuadratur}
\end{equation}
The values of $x_0$ and $m_0$
are selected such that the integrals
\begin{equation}
    I_i =\int\limits_0^\infty x^i g(x)\mathrm d x \, ,
\end{equation}
are exact for $i=0, 1$, which guarantees that Eq.~(\ref{eq:GaussQuadratur}) is exact for constant or linear functions $f(x)$. 
This gives $m_0=I_0$ and $x_0=I_1 / I_0$ in agreement with Ref.~\cite{MacDowell2019}.

Choosing the weight $g(\xi)=\alpA(\mi \xi)\alpB(\mi\xi)$
leads to the relevant mean frequency in Eq.~(\ref{eq:eff:omega})
\begin{equation}
    \overline{\omega} = \frac{\int_0^\infty \, \xi \alpA(\mi\xi)\alpB(\mi\xi)\mathrm d \xi}{\int_0^\infty \,  \alpA(\mi\xi)\alpB(\mi\xi)\mathrm d \xi}\,.\label{eq:omegabar}
\end{equation}
and Eq.~(\ref{eq:eff:omega}) directly. This equation therefore
is exact for $\varepsilon^{-2}$ constant or linear in $\mi\xi$.
We restrict ourselves to the consideration of non-retarded interactions with respect to the application of the model in density functional theory simulations. A generalisation of the model to include retardation effects is possible and reported in Ref.~\cite{MacDowell2019}.

\section{Comparison between exact and approximated $C_6$-coefficients for single oscillator models}

The simplest model for the polarizabilities $\alpha_{\rm A,B}$ is that of a single oscillator
\begin{equation}
    \alpha_{\rm A,B}(\mi\xi)=\frac{A_{\rm A,B}}{1+\left(\xi/\omega_{\rm A,B}\right)^2} \, ,\label{eq:single-oscillator}
\end{equation}
with the static value $A_{\rm A,B}$ and the resonance frequency $\omega_{\rm A,B}$. Inserting 
Eq. (\ref{eq:single-oscillator}) into Eq. (\ref{eq:omegabar})
yields the average main frequency
\begin{equation}
    \overline{\omega}=\frac{2}{\pi}\frac{\omega_{\rm A}\omega_{\rm B}}{\omega_{\rm A}-\omega_{\rm B}}\ln\left(\frac{\omega_{\rm A}}{\omega_{\rm B}}\right) \label{eq:omegabarso}
\end{equation}
giving $\overline{\omega}=2\omega_{\rm A}/\pi$ for $\omega_{\rm A}=\omega_{\rm B}$.

To illustrate the accuracy of the model assumption~(\ref{eq:eff:omega}), we calculated the averaged main frequency~(\ref{eq:omegabarso}) for different sets of resonance frequencies $\hbar\omega_{\rm A,B} \in \left[0,10\right]\,\rm{eV}$ for particles dissolved in one of the most-complex media water~\cite{Fiedler2020}.
We use the parametrization of $\varepsilon(\mi\xi)$ for water from Ref.~\cite{Fiedler2020} to define the "exact" values of the integral
\begin{align}
   C_6^{\rm exact} = \frac{3\hbar A_{\rm A}A_{\rm B}}{16\pi^3\varepsilon_0^2}\int\limits_0^\infty \frac{\mathrm d \xi}{\left[1+(\xi/\omega_{\rm A})^2\right]\left[1+(\xi/\omega_{\rm B})^2\right]\varepsilon^2(\mi\xi)}\,,
\end{align}
The comparison between the vacuum and exact Van der Waals coefficients according to Eq.~(\ref{eq:eff:omega}) allows to determine the corresponding "exact" averaged main-frequency $\omega_{\rm exact}$
\begin{equation}
    \varepsilon(\mi\omega_{\rm exact}) = \sqrt{\frac{C_6^{\rm vac}}{C_6^{\rm exact}}}\,.\label{eq:omegaexact}
\end{equation}
\begin{figure}[h]
    \centering
    \includegraphics[width=0.6\columnwidth]{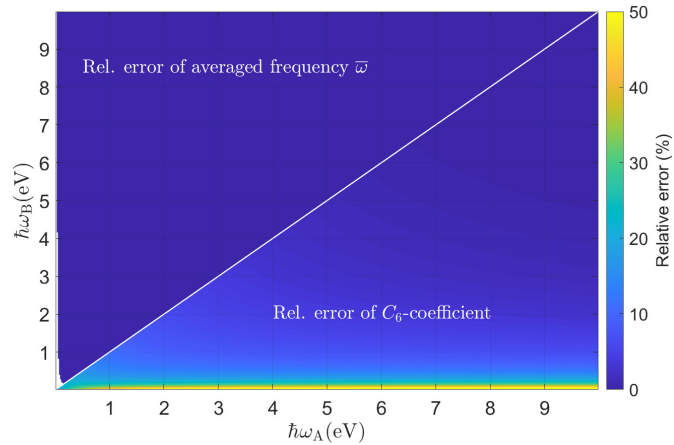}
    \caption{Relative deviations in percent between the approximated averaged frequency $\overline{\omega}$ and the exact main-frequency according to Eq.~(\ref{eq:eff:omega}) (top left triangle) and between the approximated and exact van der Waals coefficients (bottom right triangle).
}
    \label{fig:VGL_Exact}
\end{figure}
Using these values we can determine the relative deviations of the
approximated values according to Eqs.~(\ref{eq:eff:omega}) and (\ref{eq:omegabarso})
as
\begin{equation}
    \frac{\overline{\omega}-\omega_{\rm exact}}{\omega_{\rm exact}} \, ,\quad\frac{C_6^{\rm app}-C_6^{\rm exact}}{C_6^{\rm exact}}\,. \label{eq:errors}
\end{equation}

These deviations are generally rather small as depicted in Fig.~\ref{fig:VGL_Exact}. It can be observed that for materials with a dominant resonance in the microwave, optical, ultraviolet or with higher energies the relative error due to main-frequency approximation is negligible. Only for materials with a dominant resonance in the radio regime and below are not well approximated, which are not very common or realistic materials.

\section{Properties of the frequency dependent permittivity}

In order to discuss the general properties of the square of the inverse
permittivity, we consider its frequency dependence in terms of
common approximations.
The permittivity may be described in a generalised Debye form 
\begin{equation}
    \varepsilon_{\rm Debye}(\omega) =  1+\sum_D \frac{\varepsilon_D}{1-\mi\omega\tau_D}\,,
    \label{eq:genDebye}
\end{equation}
or similarly in a generalised Drude form
\begin{equation}
    \varepsilon_{\rm Drude}(\omega) = 1+\sum_D \frac{\varepsilon_D \omega_{\rm D}^2}{\omega_{\rm D}^2-\omega^2-\mi\omega\gamma_D}\,,
    \label{eq:genDrude}
\end{equation}
where the two approximations get very similar if we identify 
$\tau_D=3/\omega_D$ and $\gamma_D=3\omega_D$ (see SI). The sums in Eqs.~(\ref{eq:genDebye}) and (\ref{eq:genDrude}) contain a chosen
number of resonators. Generally, the resonator weights $\varepsilon_D$ tend to decrease with increasing resonance frequency $\omega_D$.

\begin{figure}[h]
    \centering
    \includegraphics[width=0.6\columnwidth]{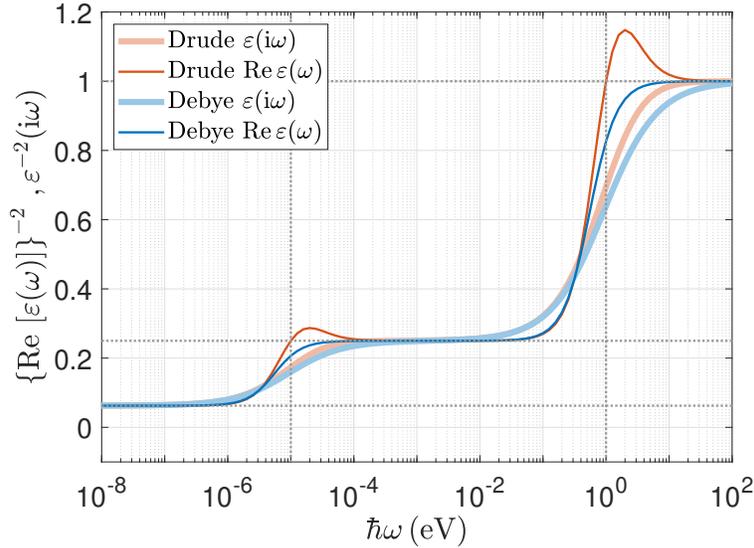}
    \caption{Drude and Debye models with two oscillators with $\hbar\omega_D=\left\lbrace10^{-5}, 1\right\rbrace \,\rm{eV}$ and weights $\varepsilon_D=\left\lbrace2, 1\right\rbrace$, respectively.
    }
    \label{fig:2resonators}
\end{figure}

Figure~\ref{fig:2resonators} shows $\varepsilon^{-2}(\mi\xi)$ in the two approximations for the model case of two resonators only. The $\varepsilon^{-2}(\mi\xi)$ is peaked around resonance frequencies, but is rather flat in other regions, where it takes the form of a step-like function. In case that $\overline{\omega}$ does not coincide with a resonance of the environment, $\varepsilon^{-2}(\mi\xi)$ can therefore be approximated by a linear function for the main part of the integral~(\ref{eq:C6}). The Gaussian quadrature is exact in this case. Figure~\ref{fig:2resonators} furthermore shows, that $\varepsilon^{-2}(\mi\xi)$ might also be replaced by 
$1/({\rm Re}[\varepsilon(\omega)])^2$ which is well measured and tabulated for many solvents and other environments~\cite{1997handbook}. Thus, the medium-assisted Van der Waals interaction can effectively be treated via a the screening due to an environmental medium
\begin{equation}
U_{\rm vdW}(d) = -\frac{C_6^{\rm vac}}{d^6} \frac{1}{\left(\rm{Re}\left[\varepsilon(\overline{\omega})\right]\right)^2} \,, 
\end{equation}
with an averaged main-frequency $\overline{\omega}$ [Eqs.~(\ref{eq:omegabarso}) and (\ref{eq:omegabarmo})]. In general the screening coefficient can be evaluated from the dielectric function $\varepsilon(\omega)$ by using the Kramers--Kronig relation~\cite{israelachvili2015intermolecular,jackson2007classical,Buhmann12a}
\begin{equation}
    \label{eq:KK}
    \varepsilon(\mi\overline{\omega}) = 1+\frac{2}{\pi}\int\limits_0^\infty \frac{\omega \,\rm{Im}\,\varepsilon(\omega)}{\omega^2+\overline{\omega}^2}\mathrm d \omega \,.
\end{equation}

\section{Impact of the model on real molecules}

In the following, we apply the model developed to more realistic scenarios and analyse the deviations between the approximation and the exact solutions for the medium-assisted van der Waals interactions. In principle, realistic materials are described via multi oscillator models, which we can distinguish into two classes according to the resonance frequencies: resonances in the ultraviolet regime, which are caused by electronic transitions, and resonances in the infrared regime that are caused by vibrational and rotational modes of the system. To this end, we first consider a two oscillator model with one oscillator within each of these spectral ranges and analyse the model predictions due the ratio between the corresponding oscillator strengths. Afterwards, we consider the interaction between real molecules whose polarizabilities consist of several oscillator models. Finally, we analyse the interaction between atomic compounds in terms of a Hamaker summation according to the common treatment of van der Waals dispersion forces in density functional theory simulations.

\subsection{Two-oscillator models}
\begin{figure}
    \centering
    \includegraphics[width=0.6\columnwidth]{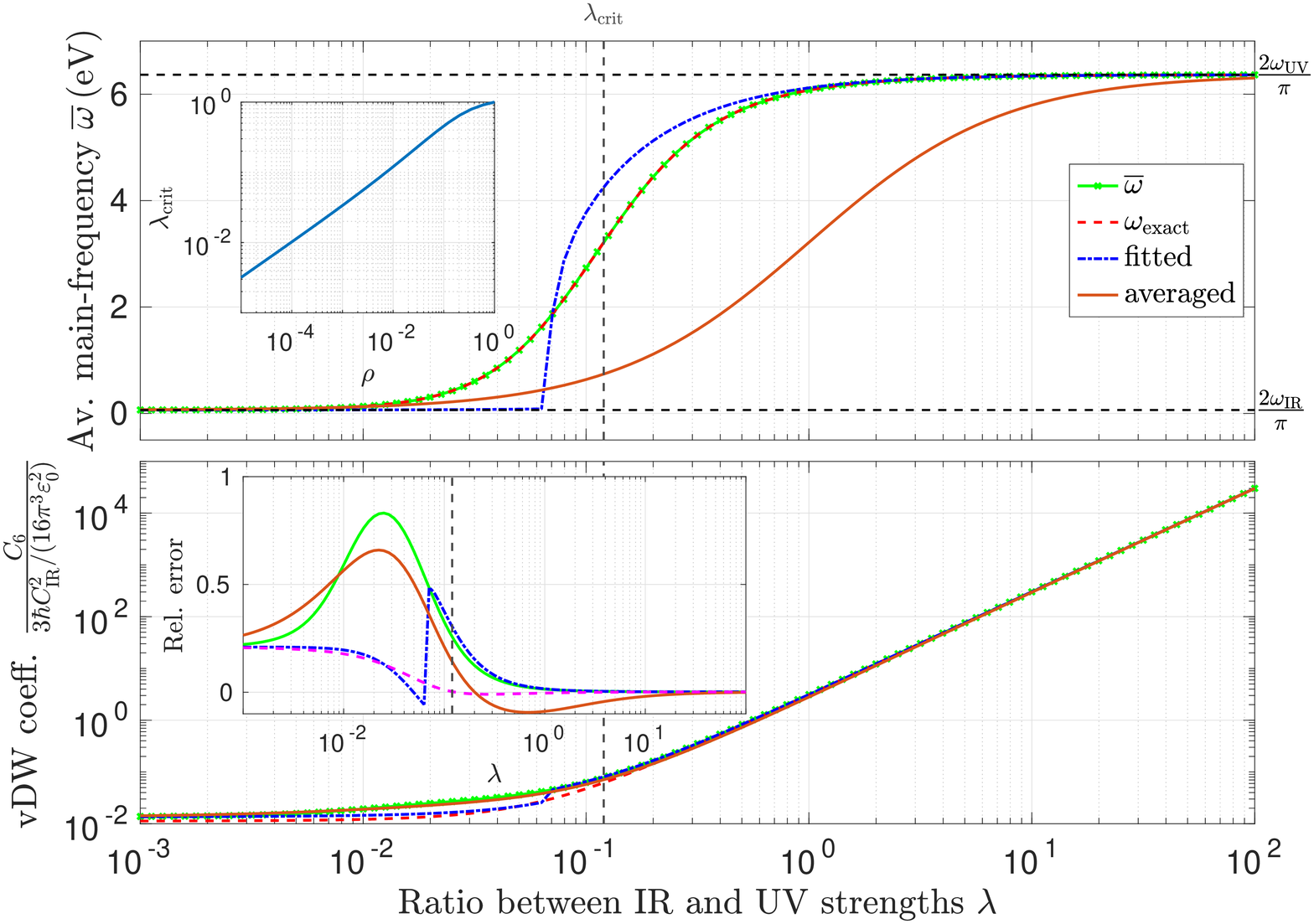}
    \caption{Comparison of the averaged main-frequency $\overline{\omega}$ (top figure) obtained via~(\ref{eq:omegabar_two}) (green lines), the exact result by solving Eq.~(\ref{eq:omegaexact}) (red lines), the fitted to a one-oscillator model~(\ref{eq:single-oscillator}) (blue lines) and an arithmetically averaged model $\overline{\omega}=2/\pi(\omega_{\rm IR}+\lambda\omega_{\rm UV})/(1+\lambda)$ (orange lines) for a two-oscillator polarizability~(\ref{eq:twoosc}) with the parameters $\hbar\omega_{\rm IR} = 0.1\,\rm{eV}$ and $\hbar\omega_{\rm UV} = 10\,\rm{eV}$ depending on the ratio between the ultraviolet and infrared oscillator strengths $\lambda$. The asymptotes of the single oscillator limits are the horizontal dashed lines. The critical ratio $\lambda_{\rm crit}$ is drawn by the vertical dashed grey line. The inset of the upper figure illustrates the dependence of the critical ratio on the ratio between the resonance frequencies $\rho=\omega_{\rm IR}/\omega_{\rm UV}$. The bottom figure illustrate the resulting normalised van der Waals coefficients with an inset of corresponding relative errors according to Eq.~(\ref{eq:errors}), where furthermore the relative error of the single oscillator treatment is depicted via the magenta line.
    }
    \label{fig:twoosc}
\end{figure}

In the previous section and also in Fig.~\ref{fig:dielectrics}, we observe that the resonances of the dielectric response function are dominant in two different spectral ranges --- in the infrared and in the ultraviolet range. Hence, we analyse the impact of differently weighted oscillator strengths in our model. 
We consider a two-oscillator model for the polarizability
\begin{align}
    \alpha(\mi\xi) = \frac{C_{\rm IR}}{1+(\xi/\omega_{\rm IR})^2} +  \frac{C_{\rm UV}}{1+(\xi/\omega_{\rm UV})^2}=C_{\rm IR}\left[\frac{1}{1+(\xi/\omega_{\rm IR})^2} +  \frac{\lambda}{1+(\xi/\omega_{\rm UV})^2}\right]\,,\label{eq:twoosc}
\end{align}
with the ratio between the oscillator strengths
\begin{equation}
    \lambda=\frac{C_{\rm UV}}{C_{\rm IR}}\,,
\end{equation}
being small ($\lambda\ll1$) for infrared-dominant species and large ($\lambda\gg1$) for ultraviolet-dominant species. By inserting Eq.~(\ref{eq:twoosc}) into Eq.~(\ref{eq:omegabar}), this leads to an averaged main-frequency
\begin{align}
    \overline{\omega} =& \frac{2}{\pi\left[\left(\omega_{\rm IR} +\lambda\omega_{\rm UV}\right)^2 +\omega_{\rm IR}\omega_{\rm UV}\left(\lambda+1\right)^2\right]}\nonumber\\
    \times&\Biggl[\frac{4\lambda\omega_{\rm IR}\omega_{\rm UV}}{\omega_{\rm IR}-\omega_{\rm UV}}\ln\left(\frac{\omega_{\rm IR}}{\omega_{\rm UV}}\right)+\left(\omega_{\rm IR}+\omega_{\rm UV}\right)\left(\lambda^2\omega_{\rm UV}^2+\omega_{\rm IR}^2\right)\Biggr]\,,\label{eq:omegabar_two}
\end{align}
satisfying the single-oscillator limits~(\ref{eq:omegabarso}).
The results for two particles of the some species with the parameters $\hbar\omega_{\rm IR} = 0.1\,\rm{eV}$ and $\hbar\omega_{\rm UV} = 10\,\rm{eV}$ embedded in water are depicted in Fig.~\ref{fig:twoosc}. It can be observed the limits of the different dominant regimes are reproduced
\begin{equation}
    \overline{\omega} = \frac{2}{\pi}\begin{cases} \omega_{\rm UV} \,, & \text{for} \,\lambda \gg \lambda_{\rm crit} \\
    \omega_{\rm IR} \,, & \text{for}\, \lambda \ll \lambda_{\rm crit} 
    \end{cases}\,.\label{eq:assym}
\end{equation}
It can be observed that $\lambda_{\rm crit}$ is smaller than unity due to the weighted integral~(\ref{eq:omegabar}). To understand this behaviour quantitatively, we define $\lambda_{\rm crit}$ to be the ratio corresponding to the arithmetical averaged main-frequency $\overline{\omega}(\lambda_{\rm crit}) = \left(\omega_{\rm IR}+\omega_{\rm UV}\right)/\pi$. This can be solved analytically and results in
\begin{align}
    \lambda_{\rm crit}&= \frac{\sqrt{\rho}}{(1+\rho)(\rho-1)^2}\left\lbrace4\rho^{3/2}\ln\rho+2\sqrt{\rho}-2\rho^{5/2}\right.\nonumber\\
    &\left.+\left[16\rho^2\ln\rho\left(\rho\ln\rho+1-\rho^2\right)+(\rho-1)^2(1+\rho)^4\right]^{1/2}\right\rbrace\,,\label{eq:lambda}
\end{align}
with the ratio between the resonance frequencies $\rho=\omega_{\rm IR}/\omega_{\rm UV}\ll1$ being typically much smaller than 1. The resulting $\lambda_{\rm crit}<1$ implies that the averaged main-frequency is typically closer to the ultraviolet resonance unless its resonance is much smaller than that of the infrared resonance.

Figure~\ref{fig:twoosc} illustrates the relevant dependencies of the two-oscillator model. In addition, to the derived two-oscillator model~(\ref{eq:omegabar_two}), we added a simple weighted averaged main-frequency model
\begin{equation}
    \overline{\omega} =\frac{2}{\pi} \frac{\omega_{\rm IR}+\lambda\omega_{\rm UV}}{1+\lambda} \,,
\end{equation}
a fit of the two-oscillator model onto a single oscillator model, which can be performed analytically by solving the corresponding least-square equation and by splitting the van der Waals coefficient into its three single oscillator contributions
\begin{align}
    \lefteqn{\frac{C_6}{3\hbar C_{\rm IR}^2/(16\pi^3\varepsilon_0^2)} = \int\limits_0^\infty \frac{\mathrm d \xi}{\varepsilon^2(\mi\xi) \left[1+\left(\xi/\omega_{\rm IR}\right)^2\right]^2}}\nonumber \\
    &+2\lambda\int\limits_0^\infty \frac{\mathrm d \xi}{\varepsilon^2(\mi\xi) \left[1+\left(\xi/\omega_{\rm IR}\right)^2\right]\cdot\left[1+\left(\xi/\omega_{\rm UV}\right)^2\right]}+\lambda^2\int\limits_0^\infty \frac{\mathrm d \xi}{\varepsilon^2(\mi\xi) \left[1+\left(\xi/\omega_{\rm UV}\right)^2\right]^2}\,. \label{eq:single_treat}
\end{align}
It can be observed, that the model~(\ref{eq:omegabar_two}) agrees very well with the exact averaged main-frequency. The  corresponding asymptotes are governed by the conditions~(\ref{eq:assym}) together with Eq.~(\ref{eq:lambda}) for $\lambda_{\rm crit}$. Remarkably, the alternative models (weighted average and fit-to-single-oscillator model) strongly deviate for the predictions of the averaged main-frequency, but matches better the van der Waals coefficients of infrared-dominant materials. For ultraviolet-dominant responding species, the presented model predicts better the $C_6$-coefficients than the other models. The optimum over the whole range is given by the treatment of each single oscillator~(\ref{eq:single_treat}).
\subsection{Multi-oscillator models}

The oscillator model~(\ref{eq:single-oscillator}) can easily be extended to several oscillators
\begin{equation}
    \alpha_{\rm A,B}(\mi\xi)=\sum_i\frac{A_{\rm A,B}^{(i)}}{1+\left(\xi/\omega_{\rm A,B}^{(i)}\right)^2} \, ,\label{eq:single-oscillatormore}
\end{equation}
leading to an averaged main-frequency
\begin{align}
\overline{\omega} = \frac{2}{\pi}\left(\sum_{i,j} A_{\rm A}^{(i)}A_{\rm B}^{(j)}\frac{\left[\omega_{\rm A}^{(i)}\omega_{\rm B}^{(j)}\right]^2}{\left[\omega_{\rm A}^{(i)}\right]^2-\left[\omega_{\rm B}^{(j)}\right]^2} \ln\left(\frac{\omega_{\rm A}^{(i)}}{\omega_{\rm B}^{(j)}}\right) \right) \left(\sum_{i,j}\frac{A_{\rm A}^{(i)}A_{\rm B}^{(j)}\omega_{\rm A}^{(i)}\omega_{\rm B}^{(j)}}{\omega_{\rm A}^{(i)}+\omega_{\rm B}^{(j)}}\right)^{-1}\,. \label{eq:omegabarmo}    
\end{align}
As a note of caution, we remark that, as evident from the discussion in Sec.~\ref{sec:two}, our method may lead to less accurate results for molecules whose relevant transitions span several plateau regions of $\varepsilon(\mi\xi)$. In this case, a separate treatment of each oscillator is recommended, as shown in the previous section. 

\begin{table}[h]
    \centering
    \begin{tabular}{c|cccccccccc}
 & CH$_4$ & NO$_2$ & CO$_2$ & CO & N$_2$O & O$_3$ & O$_2$ & N$_2$ & H$_2$S & NO \\\hline
 CH$_4$ & 11.4 & 12.6 & 12.6 & 12.2 & 12.3 & 12.7 & 13.2 & 12.8 & 10.3 & 12.8 \\
 NO$_2$ & 12.6 & 14.2 & 14.3 & 13.7 & 13.9 & 14.4 & 15.0 & 14.4 & 11.3 & 14.5 \\
 CO$_2$ & 12.6 & 14.3 & 14.3 & 13.7 & 14.0 & 14.4 & 15.0 & 14.4 & 11.4 & 14.5 \\
 CO     & 12.2 & 13.7 & 13.7 & 13.1 & 13.4 & 13.8 & 14.4 & 13.8 & 11.0 & 13.9 \\ 
 N$_2$O & 12.3 & 13.9 & 13.9 & 13.4 & 13.6 & 14.0 & 14.7 & 14.0 & 11.1 & 14.1 \\
 O$_3$  & 12.7 & 14.4 & 14.4 & 13.8 & 14.0 & 14.5 & 15.2 & 14.4 & 11.4 & 14.6 \\ 
 O$_2$  & 13.2 & 15.0 & 15.0 & 14.4 & 14.7 & 15.2 & 15.9 & 15.1 & 11.9 & 15.2 \\
 N$_2$  & 12.8 & 14.4 & 14.4 & 13.8 & 14.0 & 14.5 & 15.1 & 14.4 & 11.5 & 14.6 \\ 
 H$_2$S & 10.3 & 11.3 & 11.4 & 11.0 & 11.1 & 11.4 & 11.9 & 11.5 & 9.3 & 11.5 \\
 NO     & 12.8 & 14.5 & 14.5 & 13.9 & 14.1 & 14.6 & 15.2 & 14.6 & 11.5 & 14.7
 \end{tabular}
    \caption{Average main-frequencies $\overline{\omega} \,(\rm{eV})$ for different molecule pairs. The corresponding parameters for the polarizabilities are taken from Refs.~\cite{Johannes,Fiedler2019}.}
    \label{tab:omegabar}
\end{table}
In Table~\ref{tab:omegabar}, averaged main-frequencies are given for different pairs of small molecules. The corresponding polarizabilities are taken from Refs.~\cite{Johannes,Fiedler2019}. It can be observed that the averaged main-frequency is mainly located in the energy range between 10 and 15 eV.

\begin{figure}
    \centering
    \includegraphics[width=0.6\columnwidth]{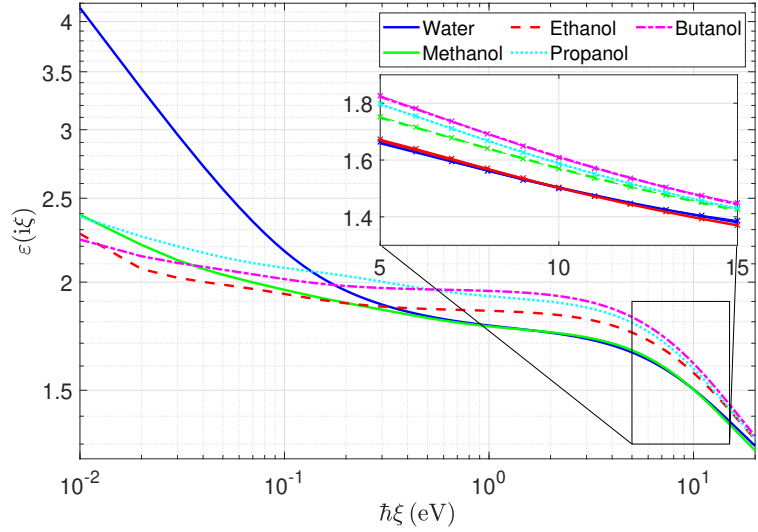}
    \caption{Dielectric function of water and some common solvent alcohols (ethanol, methanol, butanol and propanol) on the imaginary axis.The crosses in the inset illustrate the approximated dielectric functions due to Eq.~(\ref{eq:single_osc}) with the parameters in table~\ref{tab:para}.}
    \label{fig:dielectrics}
\end{figure}

Due to this reduction of the relevant energy range, the dielectric functions of the solvent can be approximated by a single UV oscillator model
\begin{equation}
    \varepsilon_{\rm app}(\mi\xi) = \frac{\varepsilon_s - \varepsilon_\infty}{1+\left(\xi/\omega_{\rm UV}\right)^2} + \varepsilon_\infty \,, \label{eq:single_osc}
\end{equation}
where $\varepsilon_s$ is the low frequency permittivity and $\varepsilon_\infty$ is the permittivity for large frequencies that may contain contributions of other oscillators at higher frequencies. The dielectric functions of typical solvents are illustrated in Fig.~\ref{fig:dielectrics}. The parametrization of the alcohols are taken from Ref.~\cite{VanZwol2010} and that of water from Ref.~\cite{Fiedler2020}. These models have been fitted to the reduced response model~(\ref{eq:single_osc}) whose resulting parameters are given in table~\ref{tab:para}. This reduced model is in good agreement with the original data which can be observed in the inset of Fig.~\ref{fig:dielectrics}. Equation~(\ref{eq:single_osc}) describes a Drude oscillator without damping, which is sufficient for the description of dispersion interactions.

\begin{table}[t]
    \centering
    \begin{tabular}{c|ccc}
    Solvent & $\varepsilon_\infty$ & $\varepsilon_s$ & $\omega_{\rm UV}$  \\\hline
    Water     & 1.193 & 1.766 & 10.73 \\
    Ethanol & 1.141 & 1.853 & 12.29 \\
    Methanol & 1.098 & 1.766 & 12.46 \\
    Butanol & 1.154 & 1.954 & 11.47 \\
    Propanol & 1.144 & 1.921 & 11.52 \\
    Cyclopentane & 1.092 & 1.938 & 11.65\\
    Cyclohexane & 1.096 & 1.991 & 11.68\\
    Benzene & 1.169 & 2.199 & 10.07\\
    Fluorobenzene & 1.145 & 2.088 & 10.31 \\
    Chlorobenzene & 1.157 & 2.264 & 10.38\\
    Bromobenzene & 1.173 & 2.371 & 10.33\\
    Pentane & 1.080 & 1.808 & 13.16\\
    Heptane & 1.086 & 1.857 & 13.07\\
    Glycerol & 1.152 & 2.152 & 12.10\\
    Tetrachloromethane & 1.103 & 2.076 & 12.13
    \end{tabular}
    \caption{Parameters for some solvents according to the reduced single oscillator model~(\ref{eq:single_osc}).}
    \label{tab:para}
\end{table}

\begin{table*}[hbt]
    \centering
    \begin{tabular}{c|c|c|c|c|c|c|c|c|c}
Mol. A & Medium & Mol. B & $C_6^{\rm vac}$ & $\overline{\omega}$ & $\varepsilon(\mi\overline{\omega})$ & $\varepsilon_{\rm app}(\mi\overline{\omega})$ & $C_6^{\rm app}$ & $C_6^{\rm exact}$ & rel. Dev. \\\hline
CH$_4$ & Water & CH$_4$ & 116.68 & 11.35 & 1.47 & 1.46 & 54.48 & 51.90 & 4.99\% \\
CO$_2$ & Water & N$_2$O & 174.42 & 13.92 & 1.40 & 1.41 & 88.16 & 81.69 & 7.91\% \\ 
NO & Water & H$_2$S & 117.12 & 11.53 & 1.46 & 1.46 & 55.02 & 52.01 & 5.80\% \\ 
CH$_4$ & Ethanol & CH$_4$ & 116.68 & 11.35 & 1.53 & 1.53 & 50.16 & 48.76 & 2.87\% \\ 
O$_2$ & Methanol & H$_2$S & 109.06 & 11.92 & 1.45 & 1.45 & 52.12 & 49.35 & 5.61\% \\
N$_2$ & Butanol & H$_2$S & 120.86 & 11.52 & 1.55 & 1.55 & 50.15 & 48.08 & 4.32\% \\ 
O$_2$ & Propanol & CH$_4$ & 81.51 & 13.24 & 1.48 & 1.48 & 37.28 & 34.74 & 7.32\% 
    \end{tabular}
    \caption{Comparison of different molecule combinations (Molecule A and B) with the vacuum Van der Waals coefficient $C_6^{\rm vac} \,(10^{79}\rm{Jm}^6)$, the corresponding averaged main-frequency $\overline{\omega} \,(\rm{eV})$, the exact and approximated~(\ref{eq:single_osc}) dielectric functions evaluated at the averaged main-frequency, the approximated Van der Waals coefficient according to Eq.~(\ref{eq:eff:omega}) $C_6^{\rm app}\, (10^{79}\rm{Jm}^6)$, the exact Van der Waals coefficient according to Eq.~(\ref{eq:C6}) $C_6^{\rm exact}\, (10^{79}\rm{Jm}^6)$ and the relative deviation of the approximated and exact $C_6$-coefficients.}
    \label{tab:impact}
\end{table*}
The impact of the model~(\ref{eq:eff:omega}) for a selection of interacting particles are given in table~\ref{tab:impact}. The complete list can be found in the supplementary information. It can be seen that the model well-approximates the Van der Waals interactions between both particles within a deviation of roughly 5-10\%. We can expect even better agreement for molecules with dominant excitations in the optical or low UV range as these frequencies are further apart from the resonances of the solvent.

\subsection{Combination model according to the summation over pairwise atomic interactions}
In common density functional theory simulations, the van der Waals interaction between complex molecules is decomposed via the Hamaker approach (pairwise summation) over the atomic constituents of each molecule~\cite{Tkatchenko2009}
\begin{equation}
    C_6^{\rm AB} = \sum_{a\in \mathcal{A}}\sum_{b\in \mathcal{B}}C_6^{ab}\,,\label{eq:C6sum}
\end{equation}
where $\mathcal{A}$ and $\mathcal{B}$ denotes the set of atoms of molecule A and B, respectively, with an effective van der Waals interaction between atom $a$ and $b$ expressed by $C_6^{ab}$.  Figure~\ref{fig:micro} illustrates this decomposition. The effective van der Waals interaction between each pair is treated analogously to the screening of the electronic density~(\ref{eq:deformation}) to be linear in the free-space van der Waals coefficient
\begin{equation}
    C_6^{ab} \propto C_6^{{\rm free},ab} \,.
\end{equation}
This ansatz allows us to apply the effective screening to the atomic decomposition of the molecules. Commonly, atomic dynamic polarizabilities are treated by a single Lorentz oscillator with an oscillator strength and a resonant frequency~(\ref{eq:single-oscillator}) that are given by the static polarizability $\alpha_{\rm A}(0)$ and the van der Waals coefficient for the equal particle pairwise interaction in vacuum
\begin{equation}
    C_6^{\rm AA} = \frac{3\hbar}{16\pi^3\varepsilon_0^2}\int\limits_0^\infty \mathrm d \xi \left[\frac{\alpha_{\rm A}(0)}{1+\left(\xi/\omega_0^{\rm A}\right)^2}\right]^2=\frac{3\hbar\alpha_{\rm A}^2(0)\omega_0^{\rm A}}{64\pi^2\varepsilon_0^2}\,.
\end{equation}
Thus, the resonance frequency is given by
\begin{equation}
    \omega_0^{\rm A} = \frac{64\pi^2\varepsilon_0^2 C_6^{\rm AA}}{3\hbar\alpha_{\rm A}^2(0)}\,.\label{eq:omega0}
\end{equation}
By inserting this~(\ref{eq:omega0}) into Eq.~(\ref{eq:omegabarso}), one finds the averaged main-frequency
\begin{equation}
    \overline{\omega}=\frac{128\pi\varepsilon_0^2}{3\hbar}\frac{C_6^{\rm AA}C_6^{\rm BB}}{C_6^{\rm AA}\alpha_{\rm B}^2(0)-C_6^{\rm BB}\alpha_{\rm A}^2(0)}\ln\left(\frac{C_6^{\rm AA}\alpha_{\rm B}^2(0)}{C_6^{\rm BB}\alpha_{\rm A}^2(0)}\right)\,,
\end{equation}
which simplifies to
\begin{equation}
    \overline{\omega}=\frac{128\pi\varepsilon_0^2C_6^{\rm AA}}{3\hbar\alpha_{\rm A}^2(0)}\,,
\end{equation}
for two particles of the same species. Table~\ref{tab:atomic_combi} illustrates the averaged main-frequency for different atomic combinations involved in organic particles: carbon, hydrogen, oxygen, nitrogen, sulfur, fluorine, chlorine, bromine and iodine. The parameters for the polarizabilities are taken from Ref.~\cite{doi:10.1063/1.1779576}. It can be seen that the resulting parameters are again in the ultraviolet range between 7.25 eV and 15 eV. Thus, the dielectric functions for the different solvents provided in table~\ref{tab:para} can be used to determine the screening factors. 
\begin{table}[htb]
    \centering
    \begin{tabular}{c|ccccccccc}
        $\hbar\overline{\omega}\,(\rm{eV})$ & C & H & O & N & S & F & Cl & Br & I  \\\hline
        C & 7.47 & 7.44 & 9.51 & 8.70 & 7.76 & 10.45 & 8.50 & 8.34 & 7.37\\H & 7.44 & 7.41 & 9.47 & 8.66 & 7.73 & 10.40 & 8.46 & 8.31 & 7.34\\O & 9.51 & 9.47 & 12.36 & 11.21 & 9.90 & 13.69 & 10.93 & 10.72 & 9.36\\N & 8.70 & 8.66 & 11.21 & 10.21 & 9.05 & 12.38 & 9.96 & 9.77 & 8.57\\S & 7.76 & 7.73 & 9.90 & 9.05 & 8.06 & 10.89 & 8.83 & 8.67 & 7.64\\F & 10.45 & 10.40 & 13.69 & 12.38 & 10.89 & 15.23 & 12.06 & 11.82 & 10.28\\Cl & 8.50 & 8.46 & 10.93 & 9.96 & 8.83 & 12.06 & 9.71 & 9.53 & 8.37\\Br & 8.34 & 8.31 & 10.72 & 9.77 & 8.67 & 11.82 & 9.53 & 9.35 & 8.22\\I & 7.37 & 7.34 & 9.36 & 8.57 & 7.64 & 10.28 & 8.37 & 8.22 & 7.26\\ 
    \end{tabular}
    \caption{The averaged main-frequency for atomic combination of organic particles.}
    \label{tab:atomic_combi}
\end{table}

The application of this atomic model has to be taken with a grain of salt because polarizabilities are typically non-additive with respect to the constituents of the considered particle~\cite{https://doi.org/10.1002/qua.24635}. There are several effects which are not covered by Eq.~(\ref{eq:C6sum}). One of these effects is the rescaling of the free-space van der Waals coefficient due to the particle's volume~(\ref{eq:deformation}). Several investigations, in theory~\cite{Ribeiro2015,D0CP02863K,doi:10.1063/1.5097553} and experiment~\cite{Coppens2019,PhysRevA.98.043403}, have shown that the largest effect of a surface (respectively a cavity) results in a spectral shift of the particle's resonance $\Delta\omega$. Such effect can, for instance, be included within this model by applying a Taylor series expansion to Eq.~(\ref{eq:omegabarso}) leading to a shift of the averaged main-frequency
\begin{align}
    \Delta\overline{\omega}&= \frac{2}{\pi}\left[\frac{\omega_{\rm A}^2\Delta\omega_{\rm B}-\omega_{\rm B}^2\Delta\omega_{\rm A}}{\left(\omega_{\rm A}-\omega_{\rm B}\right)^2}\ln\left(\frac{\omega_{\rm A}}{\omega_{\rm B}}\right) +\frac{\omega_{\rm B}\Delta\omega_{\rm A}-\omega_{rm A}\Delta\omega_{\rm B}}{\omega_{\rm A}-\omega_{\rm B}}\right.\nonumber\\
    &\left.+\frac{\omega_{\rm A}^2-\omega_{\rm B}^2 -2\omega_{\rm A}\omega_{\rm B}\ln\left(\omega_{\rm A}/\omega_{\rm B}\right)}{\left(\omega_{\rm A}-\omega_{\rm B}\right)^3}\Delta\omega_{\rm A}\Delta\omega_{\rm B}\right]\,, 
\end{align}
which has to be considered for the evaluation of the screening factor $\varepsilon^{-2}(\mi\overline{\omega}+\mi\Delta\overline{\omega})$. Another effect this model can be adapted to is the many-particle interaction behind the pairwise assumption, e.g. the three-body interaction (the Axilrod--Teller potential) that describes the interaction between three polarisable particles. Its strength is given by~\cite{doi:10.1063/1.1723844,refId0}
\begin{equation}
    C_9 \propto \int\limits_0^\infty \frac{\alpA(\mi\xi)\alpB(\mi\xi)\alpC(\mi\xi)}{\varepsilon^n(\mi\xi)}\,,
\end{equation}
where $n$ denotes the number of interactions crossing the intermediate medium, e.g. if all three particles belong to three different molecules, than $n=3$, whereas if all constituents belongs to the same molecule than $n=0$. In analogy to Eq.~(\ref{eq:omegabar}) an averaged main-frequency can be derived, which reads for three single-oscillator models
\begin{align}
    \overline{\omega}&= \frac{2}{\pi}\frac{\omega_{\rm A}\omega_{\rm B}\omega_{\rm C}}{\omega_{\rm A}+\omega_{\rm B}+\omega_{\rm C}}\frac{1}{\left(\omega_{\rm B}-\omega_{\rm C}\right)\left(\omega_{\rm A}-\omega_{\rm C}\right)\left(\omega)_{\rm A}-\omega_{\rm B}\right)}\nonumber\\
    &\times\left[\omega_{\rm A}^2\ln\left(\frac{\omega_{\rm B}}{\omega_{\rm C}}\right) +\omega_{\rm B}^2\ln\left(\frac{\omega_{\rm C}}{\omega_{\rm A}}\right)+\omega_{\rm C}^2\ln\left(\frac{\omega_{\rm A}}{\omega_{\rm C}}\right)\right]\,.
\end{align}
Beyond this extension, the model can be adapted to the consideration of higher-order multipoles, e.g. the non-retarded dipole-quadrupole interaction~\cite{Salam_2006}
\begin{equation}
    U(r)=-\frac{C_8}{r^8}\,,
\end{equation}
with
\begin{equation}
    C_8 = \frac{90\hbar c}{160\pi^3\varepsilon_0^2}\int\limits_0^\infty\md\xi\alpha(\mi\xi)\beta(\mi\xi) \,,
\end{equation}
with the scalar dipole and quadrupole polarizabilities, $\alpha(\mi\xi)$ and $\beta(\mi\xi)$, respectively. By, for instance, assuming single Lorentz oscillator models~(\ref{eq:single-oscillator}) to model each response function, the resulting screening effect can be effectively be treated in analogy to Eqs.~(\ref{eq:omegabarso}) and (\ref{eq:eff:omega}).

Beyond these extensions, there are some limitations that the model cannot cover: (i) the intermediate regime, where retardation effects start to play a role are not adaptable, because the potential does not factorises into a part depending on the polarizabilities and another part only depending on the dielectric function of the medium~\cite{PhysRevB.101.235424}; and (ii) in situations, where the interacting atomic systems A and B are so close together that the electronic densities start to overlap and a molecule is formed, the derived model fails due to the coupling dipole-electric field interaction Hamiltonian applied in the whole theory.

\section{Conclusion}

We have shown that the medium-assisted Van der Waals interaction 
can effectively be treated as a the screening due to an environmental medium. 
Single point Gauss quadrature suggests the screening to be the inverse of $\varepsilon(\mi\overline{\omega})^2$
with an averaged main-frequency $\overline{\omega}$ that depends
on the resonances of the interacting molecules. 
The approximation should be particularly accurate if these resonances are far from 
the resonances of the medium. Then $\varepsilon(\mi\overline{\omega})^{-2}$ might even be replaced by the permittivity evaluated at real frequencies $\varepsilon(\overline{\omega})^{-2}$.
Application of the approximation proposed for small molecules with resonances near to these of the solvents reveal still an accuracy of 90-95\%.

As embedded molecules in solvents often show their dominating resonances in the optical region, screening of van der Waals interactions by the solvent is greatly suppressed as compared to electrostatic Coulomb interactions. In the extreme, but important case of water, the latter are screened by the factor 1/78~\cite{Held2014} due to its large static polarizability. The permittivity in the optical region is much lower~\cite{Hartl2020}, leading to a screening of the van der Waals interaction by $\approx0.5$ compared to the vacuum case.

Nevertheless, the simple form of our result is useful for adjusting Van der Waals corrections of molecules within implicit solvents~\cite{Hartl2020}. Furthermore, the screening might affect also molecular dynamics calculations in aqueous environment such as the important problem of protein folding~\cite{dill_protein-folding_2012}, or may resolve some of the discrepancies between simulated and experimental dielectric constants~\cite{Jorge2019}. The solvents' permittivity in the optical range is caused by resonances of the solvents' electronic system. As classical force fields do not include electrons, the corresponding screening term is missing and therefore also the screening term described here.
Modification of the bare Coulomb interaction by the relative permittivity has been shown to improve the description of ion-ion interactions considerably.\cite{Kirby2019}

\bibliographystyle{unsrt}  
\bibliography{aipsamp}  





\appendix
\section{Drude and Debye formulas for a single resonator}

There are different possible forms of approximations for the frequency dependent permittivity. One of the most compact is the Debye relaxation formula for a single resonator which is completely determined by the relaxation time $\tau_D$
\begin{equation}
    \varepsilon(\omega)=\frac{\varepsilon_s-\varepsilon_\infty}{1-\mi\omega\tau_D} + \varepsilon_\infty
    \label{eq:Debye}
\end{equation}
with the static permittivity $\varepsilon_s$ and the permittivity for infinite frequency $\varepsilon_\infty$. 
This approximation reads at complex frequencies
\begin{equation}
    \varepsilon(\mi\xi)=\frac{\varepsilon_s-\varepsilon_\infty}{1+\xi\tau_D} + \varepsilon_\infty \; .
    \label{eq:iDebye}
\end{equation}
The Drude form for a single resonance at resonance frequency $\omega_D$ and width $\gamma_D$ is
\begin{equation}
    \varepsilon(\omega)=\frac{(\varepsilon_s-\varepsilon_\infty)\omega_D^2}{\omega_D^2-\omega^2-\mi\omega\gamma_D} + \varepsilon_\infty
    \label{eq:Drude}
\end{equation}
and for complex frequencies
\begin{equation}
    \varepsilon(\mi\xi)=\frac{(\varepsilon_s-\varepsilon_\infty)\omega_D^2}{\omega_D^2+\xi^2+\xi\gamma_D} + \varepsilon_\infty \; .
    \label{eq:iDrude}
\end{equation}

\begin{figure}[h!]
    \centering
    \includegraphics[width=\columnwidth]{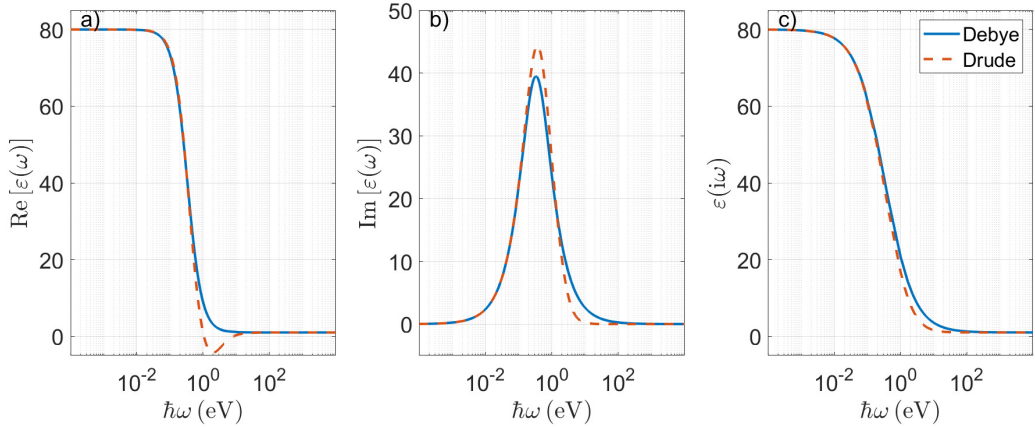}
    \caption{a) Real, b) imaginary parts of the permittivity in Debye und Drude models with $\hbar\omega_D=1\,\rm{eV},\varepsilon_s=80,\varepsilon_\infty=1$.
    c) Permittivities at imaginary frequencies. The vertical lines indicate $\omega_D$.}
    \label{fig:DDone}
\end{figure}
Drude and Debye approximations are very similar if one identifies $\tau_D=\gamma_D/\omega_D^2$ and $\gamma_D=3\omega_D$ as shown for a simple “water” model with $\hbar\omega_D=1\,\rm{eV},\varepsilon_s=80,\varepsilon_\infty=1$ in Figure \ref{fig:DDone}. 

\begin{figure}[htbp]
    \centering
    \includegraphics[width=0.6\columnwidth]{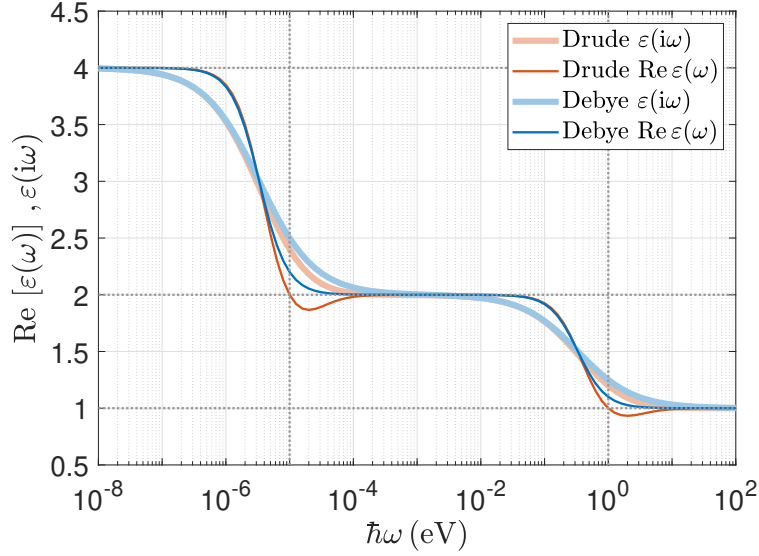}
    \caption{Comparison of the dielectric function on the imaginary frequency axis and the corresponding real part for a two oscillator model with the parameters $\hbar\omega_D=\left\lbrace10^{-5},1\right\rbrace \,\rm{eV}$ and $\varepsilon_D=\left\lbrace2,1\right\rbrace$.}
    \label{fig:DDeps}
\end{figure}
Figure~\ref{fig:DDeps} compares the real part of the dielectric function and the corresponding function at the imaginary frequency axis.

\section{Table of medium-assisted Van der Waals interaction of small molecules in different solvents}
\begin{longtable}{c|c|c|c|c|c|c|c|c|c}
\caption{Comparison of different molecule combinations (Molecule A and B) with the vacuum Van der Waals coefficient $C_6^{\rm vac} \,(10^{79}\rm{Jm}^6)$, the corresponding averaged main-frequency $\overline{\omega} \,(\rm{eV})$, the exact and approximated~(\ref{eq:single_osc}) dielectric functions evaluated at the averaged main-frequency, the approximated Van der Waals coefficient according to Eq.~(\ref{eq:eff:omega}) $C_6^{\rm app}\, (10^{79}\rm{Jm}^6)$, the exact Van der Waals coefficient according to Eq.~(\ref{eq:C6}) $C_6^{\rm exact}\, (10^{79}\rm{Jm}^6)$ and the relative deviation of the approximated and exact $C_6$-coefficients.}\\

Mol. A & Medium & Mol. B & $C_6^{\rm vac}$ & $\overline{\omega}$ & $\varepsilon(\mi\overline{\omega})$ & $\varepsilon_{\rm app}(\mi\overline{\omega})$ & $C_6^{\rm app}$ & $C_6^{\rm exact}$ & rel. Dev. \\\hline
\endfirsthead

\multicolumn{10}{c}%
{{\bfseries \tablename\ \thetable{} -- continued from previous page}} \\
Mol. A & Medium & Mol. B & $C_6^{\rm vac}$ & $\overline{\omega}$ & $\varepsilon(\mi\overline{\omega})$ & $\varepsilon_{\rm app}(\mi\overline{\omega})$ & $C_6^{\rm app}$ & $C_6^{\rm exact}$ & rel. Dev. \\\hline
\endhead

\hline \multicolumn{10}{r}{{Continued on next page}} \\ \hline
\endfoot
\hline
\endlastfoot

CH$_4$ & Water & CH$_4$ & 116.68 & 11.35 & 1.47 & 1.46 & 54.48 & 51.90 & 4.99 \\ CH$_4$ & Water & NO$_2$ & 134.21 & 12.60 & 1.43 & 1.43 & 65.27 & 61.21 & 6.63 \\ CH$_4$ & Water & CO$_2$ & 136.68 & 12.64 & 1.43 & 1.43 & 66.56 & 62.51 & 6.48 \\ CH$_4$ & Water & CO & 96.18 & 12.15 & 1.45 & 1.44 & 46.13 & 43.47 & 6.12 \\ CH$_4$ & Water & N$_2$O & 147.94 & 12.34 & 1.44 & 1.44 & 71.38 & 67.16 & 6.28 \\ CH$_4$ & Water & O$_3$ & 139.65 & 12.67 & 1.43 & 1.43 & 68.06 & 63.74 & 6.79 \\ CH$_4$ & Water & O$_2$ & 81.51 & 13.24 & 1.42 & 1.42 & 40.42 & 37.75 & 7.07 \\ CH$_4$ & Water & N$_2$ & 90.15 & 12.76 & 1.43 & 1.43 & 44.06 & 41.40 & 6.43 \\ CH$_4$ & Water & H$_2$S & 157.89 & 10.28 & 1.50 & 1.49 & 70.96 & 68.16 & 4.10 \\ CH$_4$ & Water & NO & 87.30 & 12.79 & 1.43 & 1.43 & 42.71 & 40.05 & 6.64 \\ NO$_2$ & Water & CH$_4$ & 134.21 & 12.60 & 1.43 & 1.43 & 65.27 & 61.21 & 6.63 \\ NO$_2$ & Water & NO$_2$ & 155.68 & 14.24 & 1.40 & 1.40 & 79.38 & 73.17 & 8.49 \\ NO$_2$ & Water & CO$_2$ & 158.50 & 14.26 & 1.40 & 1.40 & 80.84 & 74.66 & 8.28 \\ NO$_2$ & Water & CO & 111.21 & 13.67 & 1.41 & 1.41 & 55.82 & 51.71 & 7.95 \\ NO$_2$ & Water & N$_2$O & 171.27 & 13.90 & 1.40 & 1.41 & 86.52 & 80.03 & 8.11 \\ NO$_2$ & Water & O$_3$ & 162.11 & 14.35 & 1.39 & 1.40 & 82.90 & 76.28 & 8.67 \\ NO$_2$ & Water & O$_2$ & 94.92 & 15.03 & 1.38 & 1.39 & 49.38 & 45.37 & 8.85 \\ NO$_2$ & Water & N$_2$ & 104.61 & 14.37 & 1.39 & 1.40 & 53.51 & 49.48 & 8.16 \\ NO$_2$ & Water & H$_2$S & 180.42 & 11.34 & 1.47 & 1.46 & 84.22 & 79.64 & 5.74 \\ NO$_2$ & Water & NO & 101.36 & 14.45 & 1.39 & 1.40 & 51.97 & 47.92 & 8.44 \\ CO$_2$ & Water & CH$_4$ & 136.68 & 12.64 & 1.43 & 1.43 & 66.56 & 62.51 & 6.48 \\ CO$_2$ & Water & NO$_2$ & 158.50 & 14.26 & 1.40 & 1.40 & 80.84 & 74.66 & 8.28 \\ CO$_2$ & Water & CO$_2$ & 161.47 & 14.27 & 1.40 & 1.40 & 82.38 & 76.22 & 8.08 \\ CO$_2$ & Water & CO & 113.25 & 13.69 & 1.41 & 1.41 & 56.88 & 52.78 & 7.76 \\ CO$_2$ & Water & N$_2$O & 174.42 & 13.92 & 1.40 & 1.41 & 88.16 & 81.69 & 7.91 \\ CO$_2$ & Water & O$_3$ & 165.01 & 14.36 & 1.39 & 1.40 & 84.40 & 77.82 & 8.46 \\ CO$_2$ & Water & O$_2$ & 96.69 & 15.03 & 1.38 & 1.39 & 50.30 & 46.30 & 8.63 \\ CO$_2$ & Water & N$_2$ & 106.60 & 14.38 & 1.39 & 1.40 & 54.55 & 50.53 & 7.96 \\ CO$_2$ & Water & H$_2$S & 183.51 & 11.40 & 1.47 & 1.46 & 85.84 & 81.27 & 5.63 \\ CO$_2$ & Water & NO & 103.25 & 14.46 & 1.39 & 1.40 & 52.94 & 48.92 & 8.23 \\ CO & Water & CH$_4$ & 96.18 & 12.15 & 1.45 & 1.44 & 46.13 & 43.47 & 6.12 \\ CO & Water & NO$_2$ & 111.21 & 13.67 & 1.41 & 1.41 & 55.82 & 51.71 & 7.95 \\ CO & Water & CO$_2$ & 113.25 & 13.69 & 1.41 & 1.41 & 56.88 & 52.78 & 7.76 \\ CO & Water & CO & 79.54 & 13.13 & 1.42 & 1.42 & 39.32 & 36.61 & 7.41 \\ CO & Water & N$_2$O & 122.44 & 13.35 & 1.42 & 1.42 & 60.91 & 56.62 & 7.57 \\ CO & Water & O$_3$ & 115.77 & 13.76 & 1.41 & 1.41 & 58.26 & 53.88 & 8.13 \\ CO & Water & O$_2$ & 67.71 & 14.41 & 1.39 & 1.40 & 34.67 & 32.00 & 8.36 \\ CO & Water & N$_2$ & 74.72 & 13.80 & 1.41 & 1.41 & 37.65 & 34.97 & 7.66 \\ CO & Water & H$_2$S & 129.60 & 10.96 & 1.48 & 1.47 & 59.71 & 56.75 & 5.21 \\ CO & Water & NO & 72.38 & 13.87 & 1.41 & 1.41 & 36.54 & 33.85 & 7.93 \\ N$_2$O & Water & CH$_4$ & 147.94 & 12.34 & 1.44 & 1.44 & 71.38 & 67.16 & 6.28 \\ N$_2$O & Water & NO$_2$ & 171.27 & 13.90 & 1.40 & 1.41 & 86.52 & 80.03 & 8.11 \\ N$_2$O & Water & CO$_2$ & 174.42 & 13.92 & 1.40 & 1.41 & 88.16 & 81.69 & 7.91 \\ N$_2$O & Water & CO & 122.44 & 13.35 & 1.42 & 1.42 & 60.91 & 56.62 & 7.57 \\ N$_2$O & Water & N$_2$O & 188.52 & 13.58 & 1.41 & 1.41 & 94.38 & 87.60 & 7.73 \\ N$_2$O & Water & O$_3$ & 178.30 & 14.00 & 1.40 & 1.41 & 90.32 & 83.41 & 8.29 \\ N$_2$O & Water & O$_2$ & 104.35 & 14.65 & 1.39 & 1.39 & 53.78 & 49.57 & 8.49 \\ N$_2$O & Water & N$_2$ & 115.11 & 14.03 & 1.40 & 1.40 & 58.36 & 54.13 & 7.80 \\ N$_2$O & Water & H$_2$S & 199.10 & 11.13 & 1.47 & 1.47 & 92.26 & 87.54 & 5.39 \\ N$_2$O & Water & NO & 111.50 & 14.10 & 1.40 & 1.40 & 56.64 & 52.41 & 8.07 \\ O$_3$ & Water & CH$_4$ & 139.65 & 12.67 & 1.43 & 1.43 & 68.06 & 63.74 & 6.79 \\ O$_3$ & Water & NO$_2$ & 162.11 & 14.35 & 1.39 & 1.40 & 82.90 & 76.28 & 8.67 \\ O$_3$ & Water & CO$_2$ & 165.01 & 14.36 & 1.39 & 1.40 & 84.40 & 77.82 & 8.46 \\ O$_3$ & Water & CO & 115.77 & 13.76 & 1.41 & 1.41 & 58.26 & 53.88 & 8.13 \\ O$_3$ & Water & N$_2$O & 178.30 & 14.00 & 1.40 & 1.41 & 90.32 & 83.41 & 8.29 \\ O$_3$ & Water & O$_3$ & 168.83 & 14.47 & 1.39 & 1.40 & 86.59 & 79.54 & 8.86 \\ O$_3$ & Water & O$_2$ & 98.84 & 15.16 & 1.38 & 1.38 & 51.58 & 47.31 & 9.02 \\ O$_3$ & Water & N$_2$ & 108.90 & 14.48 & 1.39 & 1.40 & 55.87 & 51.57 & 8.33 \\ O$_3$ & Water & H$_2$S & 187.75 & 11.39 & 1.47 & 1.46 & 87.79 & 82.90 & 5.89 \\ O$_3$ & Water & NO & 105.53 & 14.56 & 1.39 & 1.39 & 54.26 & 49.96 & 8.62 \\ O$_2$ & Water & CH$_4$ & 81.51 & 13.24 & 1.42 & 1.42 & 40.42 & 37.75 & 7.07 \\ O$_2$ & Water & NO$_2$ & 94.92 & 15.03 & 1.38 & 1.39 & 49.38 & 45.37 & 8.85 \\ O$_2$ & Water & CO$_2$ & 96.69 & 15.03 & 1.38 & 1.39 & 50.30 & 46.30 & 8.63 \\ O$_2$ & Water & CO & 67.71 & 14.41 & 1.39 & 1.40 & 34.67 & 32.00 & 8.36 \\ O$_2$ & Water & N$_2$O & 104.35 & 14.65 & 1.39 & 1.39 & 53.78 & 49.57 & 8.49 \\ O$_2$ & Water & O$_3$ & 98.84 & 15.16 & 1.38 & 1.38 & 51.58 & 47.31 & 9.02 \\ O$_2$ & Water & O$_2$ & 58.01 & 15.86 & 1.36 & 1.37 & 30.78 & 28.20 & 9.12 \\ O$_2$ & Water & N$_2$ & 63.86 & 15.13 & 1.38 & 1.38 & 33.30 & 30.70 & 8.48 \\ O$_2$ & Water & H$_2$S & 109.06 & 11.92 & 1.45 & 1.45 & 51.92 & 48.85 & 6.27 \\ O$_2$ & Water & NO & 61.86 & 15.24 & 1.38 & 1.38 & 32.34 & 29.74 & 8.77 \\ N$_2$ & Water & CH$_4$ & 90.15 & 12.76 & 1.43 & 1.43 & 44.06 & 41.40 & 6.43 \\ N$_2$ & Water & NO$_2$ & 104.61 & 14.37 & 1.39 & 1.40 & 53.51 & 49.48 & 8.16 \\ N$_2$ & Water & CO$_2$ & 106.60 & 14.38 & 1.39 & 1.40 & 54.55 & 50.53 & 7.96 \\ N$_2$ & Water & CO & 74.72 & 13.80 & 1.41 & 1.41 & 37.65 & 34.97 & 7.66 \\ N$_2$ & Water & N$_2$O & 115.11 & 14.03 & 1.40 & 1.40 & 58.36 & 54.13 & 7.80 \\ N$_2$ & Water & O$_3$ & 108.90 & 14.48 & 1.39 & 1.40 & 55.87 & 51.57 & 8.33 \\ N$_2$ & Water & O$_2$ & 63.86 & 15.13 & 1.38 & 1.38 & 33.30 & 30.70 & 8.48 \\ N$_2$ & Water & N$_2$ & 70.40 & 14.49 & 1.39 & 1.40 & 36.12 & 33.50 & 7.83 \\ N$_2$ & Water & H$_2$S & 120.86 & 11.52 & 1.46 & 1.46 & 56.76 & 53.74 & 5.61 \\ N$_2$ & Water & NO & 68.17 & 14.57 & 1.39 & 1.39 & 35.05 & 32.43 & 8.10 \\ H$_2$S & Water & CH$_4$ & 157.89 & 10.28 & 1.50 & 1.49 & 70.96 & 68.16 & 4.10 \\ H$_2$S & Water & NO$_2$ & 180.42 & 11.34 & 1.47 & 1.46 & 84.22 & 79.64 & 5.74 \\ H$_2$S & Water & CO$_2$ & 183.51 & 11.40 & 1.47 & 1.46 & 85.84 & 81.27 & 5.63 \\ H$_2$S & Water & CO & 129.60 & 10.96 & 1.48 & 1.47 & 59.71 & 56.75 & 5.21 \\ H$_2$S & Water & N$_2$O & 199.10 & 11.13 & 1.47 & 1.47 & 92.26 & 87.54 & 5.39 \\ H$_2$S & Water & O$_3$ & 187.75 & 11.39 & 1.47 & 1.46 & 87.79 & 82.90 & 5.89 \\ H$_2$S & Water & O$_2$ & 109.06 & 11.92 & 1.45 & 1.45 & 51.92 & 48.85 & 6.27 \\ H$_2$S & Water & N$_2$ & 120.86 & 11.52 & 1.46 & 1.46 & 56.76 & 53.74 & 5.61 \\ H$_2$S & Water & H$_2$S & 215.59 & 9.33 & 1.52 & 1.52 & 93.40 & 90.51 & 3.19 \\ H$_2$S & Water & NO & 117.12 & 11.53 & 1.46 & 1.46 & 55.02 & 52.01 & 5.80 \\ NO & Water & CH$_4$ & 87.30 & 12.79 & 1.43 & 1.43 & 42.71 & 40.05 & 6.64 \\ NO & Water & NO$_2$ & 101.36 & 14.45 & 1.39 & 1.40 & 51.97 & 47.92 & 8.44 \\ NO & Water & CO$_2$ & 103.25 & 14.46 & 1.39 & 1.40 & 52.94 & 48.92 & 8.23 \\ NO & Water & CO & 72.38 & 13.87 & 1.41 & 1.41 & 36.54 & 33.85 & 7.93 \\ NO & Water & N$_2$O & 111.50 & 14.10 & 1.40 & 1.40 & 56.64 & 52.41 & 8.07 \\ NO & Water & O$_3$ & 105.53 & 14.56 & 1.39 & 1.39 & 54.26 & 49.96 & 8.62 \\ NO & Water & O$_2$ & 61.86 & 15.24 & 1.38 & 1.38 & 32.34 & 29.74 & 8.77 \\ NO & Water & N$_2$ & 68.17 & 14.57 & 1.39 & 1.39 & 35.05 & 32.43 & 8.10 \\ NO & Water & H$_2$S & 117.12 & 11.53 & 1.46 & 1.46 & 55.02 & 52.01 & 5.80 \\ NO & Water & NO & 66.03 & 14.66 & 1.39 & 1.39 & 34.03 & 31.40 & 8.38 \\  
    
CH$_4$ & Ethanol & CH$_4$ & 116.68 & 11.35 & 1.53 & 1.53 & 50.16 & 48.76 & 2.87 \\ CH$_4$ & Ethanol & NO$_2$ & 134.21 & 12.60 & 1.49 & 1.49 & 60.59 & 57.71 & 4.99 \\ CH$_4$ & Ethanol & CO$_2$ & 136.68 & 12.64 & 1.49 & 1.49 & 61.81 & 58.93 & 4.89 \\ CH$_4$ & Ethanol & CO & 96.18 & 12.15 & 1.50 & 1.50 & 42.69 & 40.93 & 4.30 \\ CH$_4$ & Ethanol & N$_2$O & 147.94 & 12.34 & 1.50 & 1.50 & 66.15 & 63.27 & 4.55 \\ CH$_4$ & Ethanol & O$_3$ & 139.65 & 12.67 & 1.49 & 1.49 & 63.22 & 60.12 & 5.16 \\ CH$_4$ & Ethanol & O$_2$ & 81.51 & 13.24 & 1.47 & 1.47 & 37.70 & 35.65 & 5.75 \\ CH$_4$ & Ethanol & N$_2$ & 90.15 & 12.76 & 1.48 & 1.48 & 40.95 & 39.04 & 4.89 \\ CH$_4$ & Ethanol & H$_2$S & 157.89 & 10.28 & 1.56 & 1.56 & 64.90 & 63.85 & 1.64 \\ CH$_4$ & Ethanol & NO & 87.30 & 12.79 & 1.48 & 1.48 & 39.70 & 37.77 & 5.11 \\ NO$_2$ & Ethanol & CH$_4$ & 134.21 & 12.60 & 1.49 & 1.49 & 60.59 & 57.71 & 4.99 \\ NO$_2$ & Ethanol & NO$_2$ & 155.68 & 14.24 & 1.44 & 1.44 & 74.57 & 69.28 & 7.64 \\ NO$_2$ & Ethanol & CO$_2$ & 158.50 & 14.26 & 1.44 & 1.44 & 75.96 & 70.67 & 7.48 \\ NO$_2$ & Ethanol & CO & 111.21 & 13.67 & 1.46 & 1.46 & 52.22 & 48.88 & 6.83 \\ NO$_2$ & Ethanol & N$_2$O & 171.27 & 13.90 & 1.45 & 1.45 & 81.08 & 75.70 & 7.11 \\ NO$_2$ & Ethanol & O$_3$ & 162.11 & 14.35 & 1.44 & 1.44 & 77.94 & 72.26 & 7.86 \\ NO$_2$ & Ethanol & O$_2$ & 94.92 & 15.03 & 1.42 & 1.43 & 46.66 & 43.03 & 8.44 \\ NO$_2$ & Ethanol & N$_2$ & 104.61 & 14.37 & 1.44 & 1.44 & 50.32 & 46.85 & 7.41 \\ NO$_2$ & Ethanol & H$_2$S & 180.42 & 11.34 & 1.53 & 1.53 & 77.52 & 74.85 & 3.57 \\ NO$_2$ & Ethanol & NO & 101.36 & 14.45 & 1.44 & 1.44 & 48.89 & 45.39 & 7.72 \\ CO$_2$ & Ethanol & CH$_4$ & 136.68 & 12.64 & 1.49 & 1.49 & 61.81 & 58.93 & 4.89 \\ CO$_2$ & Ethanol & NO$_2$ & 158.50 & 14.26 & 1.44 & 1.44 & 75.96 & 70.67 & 7.48 \\ CO$_2$ & Ethanol & CO$_2$ & 161.47 & 14.27 & 1.44 & 1.44 & 77.41 & 72.13 & 7.31 \\ CO$_2$ & Ethanol & CO & 113.25 & 13.69 & 1.46 & 1.46 & 53.22 & 49.88 & 6.69 \\ CO$_2$ & Ethanol & N$_2$O & 174.42 & 13.92 & 1.45 & 1.45 & 82.62 & 77.25 & 6.95 \\ CO$_2$ & Ethanol & O$_3$ & 165.01 & 14.36 & 1.44 & 1.44 & 79.36 & 73.69 & 7.69 \\ CO$_2$ & Ethanol & O$_2$ & 96.69 & 15.03 & 1.42 & 1.43 & 47.52 & 43.90 & 8.25 \\ CO$_2$ & Ethanol & N$_2$ & 106.60 & 14.38 & 1.44 & 1.44 & 51.29 & 47.83 & 7.24 \\ CO$_2$ & Ethanol & H$_2$S & 183.51 & 11.40 & 1.53 & 1.52 & 79.05 & 76.36 & 3.52 \\ CO$_2$ & Ethanol & NO & 103.25 & 14.46 & 1.44 & 1.44 & 49.82 & 46.32 & 7.55 \\ CO & Ethanol & CH$_4$ & 96.18 & 12.15 & 1.50 & 1.50 & 42.69 & 40.93 & 4.30 \\ CO & Ethanol & NO$_2$ & 111.21 & 13.67 & 1.46 & 1.46 & 52.22 & 48.88 & 6.83 \\ CO & Ethanol & CO$_2$ & 113.25 & 13.69 & 1.46 & 1.46 & 53.22 & 49.88 & 6.69 \\ CO & Ethanol & CO & 79.54 & 13.13 & 1.47 & 1.47 & 36.64 & 34.55 & 6.04 \\ CO & Ethanol & N$_2$O & 122.44 & 13.35 & 1.47 & 1.47 & 56.85 & 53.47 & 6.32 \\ CO & Ethanol & O$_3$ & 115.77 & 13.76 & 1.46 & 1.46 & 54.55 & 50.96 & 7.04 \\ CO & Ethanol & O$_2$ & 67.71 & 14.41 & 1.44 & 1.44 & 32.61 & 30.30 & 7.64 \\ CO & Ethanol & N$_2$ & 74.72 & 13.80 & 1.45 & 1.46 & 35.25 & 33.06 & 6.64 \\ CO & Ethanol & H$_2$S & 129.60 & 10.96 & 1.54 & 1.54 & 54.83 & 53.27 & 2.93 \\ CO & Ethanol & NO & 72.38 & 13.87 & 1.45 & 1.45 & 34.23 & 32.01 & 6.93 \\ N$_2$O & Ethanol & CH$_4$ & 147.94 & 12.34 & 1.50 & 1.50 & 66.15 & 63.27 & 4.55 \\ N$_2$O & Ethanol & NO$_2$ & 171.27 & 13.90 & 1.45 & 1.45 & 81.08 & 75.70 & 7.11 \\ N$_2$O & Ethanol & CO$_2$ & 174.42 & 13.92 & 1.45 & 1.45 & 82.62 & 77.25 & 6.95 \\ N$_2$O & Ethanol & CO & 122.44 & 13.35 & 1.47 & 1.47 & 56.85 & 53.47 & 6.32 \\ N$_2$O & Ethanol & N$_2$O & 188.52 & 13.58 & 1.46 & 1.46 & 88.24 & 82.78 & 6.59 \\ N$_2$O & Ethanol & O$_3$ & 178.30 & 14.00 & 1.45 & 1.45 & 84.71 & 78.93 & 7.32 \\ N$_2$O & Ethanol & O$_2$ & 104.35 & 14.65 & 1.43 & 1.44 & 50.67 & 46.96 & 7.90 \\ N$_2$O & Ethanol & N$_2$ & 115.11 & 14.03 & 1.45 & 1.45 & 54.74 & 51.21 & 6.90 \\ N$_2$O & Ethanol & H$_2$S & 199.10 & 11.13 & 1.53 & 1.53 & 84.81 & 82.21 & 3.17 \\ N$_2$O & Ethanol & NO & 111.50 & 14.10 & 1.45 & 1.45 & 53.16 & 49.59 & 7.19 \\ O$_3$ & Ethanol & CH$_4$ & 139.65 & 12.67 & 1.49 & 1.49 & 63.22 & 60.12 & 5.16 \\ O$_3$ & Ethanol & NO$_2$ & 162.11 & 14.35 & 1.44 & 1.44 & 77.94 & 72.26 & 7.86 \\ O$_3$ & Ethanol & CO$_2$ & 165.01 & 14.36 & 1.44 & 1.44 & 79.36 & 73.69 & 7.69 \\ O$_3$ & Ethanol & CO & 115.77 & 13.76 & 1.46 & 1.46 & 54.55 & 50.96 & 7.04 \\ O$_3$ & Ethanol & N$_2$O & 178.30 & 14.00 & 1.45 & 1.45 & 84.71 & 78.93 & 7.32 \\ O$_3$ & Ethanol & O$_3$ & 168.83 & 14.47 & 1.44 & 1.44 & 81.48 & 75.38 & 8.09 \\ O$_3$ & Ethanol & O$_2$ & 98.84 & 15.16 & 1.42 & 1.42 & 48.78 & 44.89 & 8.67 \\ O$_3$ & Ethanol & N$_2$ & 108.90 & 14.48 & 1.44 & 1.44 & 52.57 & 48.85 & 7.62 \\ O$_3$ & Ethanol & H$_2$S & 187.75 & 11.39 & 1.53 & 1.52 & 80.83 & 77.94 & 3.71 \\ O$_3$ & Ethanol & NO & 105.53 & 14.56 & 1.43 & 1.44 & 51.10 & 47.34 & 7.94 \\ O$_2$ & Ethanol & CH$_4$ & 81.51 & 13.24 & 1.47 & 1.47 & 37.70 & 35.65 & 5.75 \\ O$_2$ & Ethanol & NO$_2$ & 94.92 & 15.03 & 1.42 & 1.43 & 46.66 & 43.03 & 8.44 \\ O$_2$ & Ethanol & CO$_2$ & 96.69 & 15.03 & 1.42 & 1.43 & 47.52 & 43.90 & 8.25 \\ O$_2$ & Ethanol & CO & 67.71 & 14.41 & 1.44 & 1.44 & 32.61 & 30.30 & 7.64 \\ O$_2$ & Ethanol & N$_2$O & 104.35 & 14.65 & 1.43 & 1.44 & 50.67 & 46.96 & 7.90 \\ O$_2$ & Ethanol & O$_3$ & 98.84 & 15.16 & 1.42 & 1.42 & 48.78 & 44.89 & 8.67 \\ O$_2$ & Ethanol & O$_2$ & 58.01 & 15.86 & 1.40 & 1.41 & 29.26 & 26.80 & 9.18 \\ O$_2$ & Ethanol & N$_2$ & 63.86 & 15.13 & 1.42 & 1.42 & 31.49 & 29.12 & 8.15 \\ O$_2$ & Ethanol & H$_2$S & 109.06 & 11.92 & 1.51 & 1.51 & 47.97 & 45.97 & 4.35 \\ O$_2$ & Ethanol & NO & 61.86 & 15.24 & 1.42 & 1.42 & 30.61 & 28.21 & 8.49 \\ N$_2$ & Ethanol & CH$_4$ & 90.15 & 12.76 & 1.48 & 1.48 & 40.95 & 39.04 & 4.89 \\ N$_2$ & Ethanol & NO$_2$ & 104.61 & 14.37 & 1.44 & 1.44 & 50.32 & 46.85 & 7.41 \\ N$_2$ & Ethanol & CO$_2$ & 106.60 & 14.38 & 1.44 & 1.44 & 51.29 & 47.83 & 7.24 \\ N$_2$ & Ethanol & CO & 74.72 & 13.80 & 1.45 & 1.46 & 35.25 & 33.06 & 6.64 \\ N$_2$ & Ethanol & N$_2$O & 115.11 & 14.03 & 1.45 & 1.45 & 54.74 & 51.21 & 6.90 \\ N$_2$ & Ethanol & O$_3$ & 108.90 & 14.48 & 1.44 & 1.44 & 52.57 & 48.85 & 7.62 \\ N$_2$ & Ethanol & O$_2$ & 63.86 & 15.13 & 1.42 & 1.42 & 31.49 & 29.12 & 8.15 \\ N$_2$ & Ethanol & N$_2$ & 70.40 & 14.49 & 1.44 & 1.44 & 34.00 & 31.72 & 7.17 \\ N$_2$ & Ethanol & H$_2$S & 120.86 & 11.52 & 1.52 & 1.52 & 52.30 & 50.51 & 3.55 \\ N$_2$ & Ethanol & NO & 68.17 & 14.57 & 1.43 & 1.44 & 33.01 & 30.72 & 7.47 \\ H$_2$S & Ethanol & CH$_4$ & 157.89 & 10.28 & 1.56 & 1.56 & 64.90 & 63.85 & 1.64 \\ H$_2$S & Ethanol & NO$_2$ & 180.42 & 11.34 & 1.53 & 1.53 & 77.52 & 74.85 & 3.57 \\ H$_2$S & Ethanol & CO$_2$ & 183.51 & 11.40 & 1.53 & 1.52 & 79.05 & 76.36 & 3.52 \\ H$_2$S & Ethanol & CO & 129.60 & 10.96 & 1.54 & 1.54 & 54.83 & 53.27 & 2.93 \\ H$_2$S & Ethanol & N$_2$O & 199.10 & 11.13 & 1.53 & 1.53 & 84.81 & 82.21 & 3.17 \\ H$_2$S & Ethanol & O$_3$ & 187.75 & 11.39 & 1.53 & 1.52 & 80.83 & 77.94 & 3.71 \\ H$_2$S & Ethanol & O$_2$ & 109.06 & 11.92 & 1.51 & 1.51 & 47.97 & 45.97 & 4.35 \\ H$_2$S & Ethanol & N$_2$ & 120.86 & 11.52 & 1.52 & 1.52 & 52.30 & 50.51 & 3.55 \\ H$_2$S & Ethanol & H$_2$S & 215.59 & 9.33 & 1.59 & 1.59 & 84.99 & 84.56 & 0.50 \\ H$_2$S & Ethanol & NO & 117.12 & 11.53 & 1.52 & 1.52 & 50.71 & 48.89 & 3.72 \\ NO & Ethanol & CH$_4$ & 87.30 & 12.79 & 1.48 & 1.48 & 39.70 & 37.77 & 5.11 \\ NO & Ethanol & NO$_2$ & 101.36 & 14.45 & 1.44 & 1.44 & 48.89 & 45.39 & 7.72 \\ NO & Ethanol & CO$_2$ & 103.25 & 14.46 & 1.44 & 1.44 & 49.82 & 46.32 & 7.55 \\ NO & Ethanol & CO & 72.38 & 13.87 & 1.45 & 1.45 & 34.23 & 32.01 & 6.93 \\ NO & Ethanol & N$_2$O & 111.50 & 14.10 & 1.45 & 1.45 & 53.16 & 49.59 & 7.19 \\ NO & Ethanol & O$_3$ & 105.53 & 14.56 & 1.43 & 1.44 & 51.10 & 47.34 & 7.94 \\ NO & Ethanol & O$_2$ & 61.86 & 15.24 & 1.42 & 1.42 & 30.61 & 28.21 & 8.49 \\ NO & Ethanol & N$_2$ & 68.17 & 14.57 & 1.43 & 1.44 & 33.01 & 30.72 & 7.47 \\ NO & Ethanol & H$_2$S & 117.12 & 11.53 & 1.52 & 1.52 & 50.71 & 48.89 & 3.72 \\ NO & Ethanol & NO & 66.03 & 14.66 & 1.43 & 1.43 & 32.07 & 29.75 & 7.79 \\ 

CH$_4$ & Methanol & CH$_4$ & 116.68 & 11.35 & 1.46 & 1.46 & 54.52 & 52.41 & 4.04 \\ CH$_4$ & Methanol & NO$_2$ & 134.21 & 12.60 & 1.43 & 1.43 & 65.79 & 61.87 & 6.33 \\ CH$_4$ & Methanol & CO$_2$ & 136.68 & 12.64 & 1.43 & 1.43 & 67.11 & 63.18 & 6.22 \\ CH$_4$ & Methanol & CO & 96.18 & 12.15 & 1.44 & 1.44 & 46.37 & 43.92 & 5.58 \\ CH$_4$ & Methanol & N$_2$O & 147.94 & 12.34 & 1.43 & 1.44 & 71.84 & 67.87 & 5.85 \\ CH$_4$ & Methanol & O$_3$ & 139.65 & 12.67 & 1.43 & 1.43 & 68.64 & 64.44 & 6.52 \\ CH$_4$ & Methanol & O$_2$ & 81.51 & 13.24 & 1.41 & 1.41 & 40.91 & 38.18 & 7.16 \\ CH$_4$ & Methanol & N$_2$ & 90.15 & 12.76 & 1.42 & 1.42 & 44.45 & 41.85 & 6.22 \\ CH$_4$ & Methanol & H$_2$S & 157.89 & 10.28 & 1.49 & 1.50 & 70.63 & 68.76 & 2.71 \\ CH$_4$ & Methanol & NO & 87.30 & 12.79 & 1.42 & 1.42 & 43.10 & 40.49 & 6.46 \\ NO$_2$ & Methanol & CH$_4$ & 134.21 & 12.60 & 1.43 & 1.43 & 65.79 & 61.87 & 6.33 \\ NO$_2$ & Methanol & NO$_2$ & 155.68 & 14.24 & 1.39 & 1.39 & 80.87 & 74.04 & 9.22 \\ NO$_2$ & Methanol & CO$_2$ & 158.50 & 14.26 & 1.39 & 1.39 & 82.37 & 75.55 & 9.03 \\ NO$_2$ & Methanol & CO & 111.21 & 13.67 & 1.40 & 1.40 & 56.65 & 52.30 & 8.32 \\ NO$_2$ & Methanol & N$_2$O & 171.27 & 13.90 & 1.40 & 1.40 & 87.95 & 80.97 & 8.62 \\ NO$_2$ & Methanol & O$_3$ & 162.11 & 14.35 & 1.38 & 1.38 & 84.52 & 77.20 & 9.47 \\ NO$_2$ & Methanol & O$_2$ & 94.92 & 15.03 & 1.37 & 1.37 & 50.57 & 45.93 & 10.11 \\ NO$_2$ & Methanol & N$_2$ & 104.61 & 14.37 & 1.38 & 1.38 & 54.57 & 50.07 & 8.97 \\ NO$_2$ & Methanol & H$_2$S & 180.42 & 11.34 & 1.46 & 1.46 & 84.27 & 80.42 & 4.78 \\ NO$_2$ & Methanol & NO & 101.36 & 14.45 & 1.38 & 1.38 & 53.02 & 48.50 & 9.31 \\ CO$_2$ & Methanol & CH$_4$ & 136.68 & 12.64 & 1.43 & 1.43 & 67.11 & 63.18 & 6.22 \\ CO$_2$ & Methanol & NO$_2$ & 158.50 & 14.26 & 1.39 & 1.39 & 82.37 & 75.55 & 9.03 \\ CO$_2$ & Methanol & CO$_2$ & 161.47 & 14.27 & 1.39 & 1.39 & 83.94 & 77.13 & 8.84 \\ CO$_2$ & Methanol & CO & 113.25 & 13.69 & 1.40 & 1.40 & 57.73 & 53.38 & 8.15 \\ CO$_2$ & Methanol & N$_2$O & 174.42 & 13.92 & 1.40 & 1.40 & 89.62 & 82.64 & 8.45 \\ CO$_2$ & Methanol & O$_3$ & 165.01 & 14.36 & 1.38 & 1.38 & 86.06 & 78.75 & 9.28 \\ CO$_2$ & Methanol & O$_2$ & 96.69 & 15.03 & 1.37 & 1.37 & 51.51 & 46.88 & 9.89 \\ CO$_2$ & Methanol & N$_2$ & 106.60 & 14.38 & 1.38 & 1.38 & 55.62 & 51.13 & 8.78 \\ CO$_2$ & Methanol & H$_2$S & 183.51 & 11.40 & 1.46 & 1.46 & 85.93 & 82.06 & 4.71 \\ CO$_2$ & Methanol & NO & 103.25 & 14.46 & 1.38 & 1.38 & 54.02 & 49.50 & 9.11 \\ CO & Methanol & CH$_4$ & 96.18 & 12.15 & 1.44 & 1.44 & 46.37 & 43.92 & 5.58 \\ CO & Methanol & NO$_2$ & 111.21 & 13.67 & 1.40 & 1.40 & 56.65 & 52.30 & 8.32 \\ CO & Methanol & CO$_2$ & 113.25 & 13.69 & 1.40 & 1.40 & 57.73 & 53.38 & 8.15 \\ CO & Methanol & CO & 79.54 & 13.13 & 1.41 & 1.41 & 39.77 & 37.01 & 7.45 \\ CO & Methanol & N$_2$O & 122.44 & 13.35 & 1.41 & 1.41 & 61.69 & 57.26 & 7.75 \\ CO & Methanol & O$_3$ & 115.77 & 13.76 & 1.40 & 1.40 & 59.17 & 54.51 & 8.55 \\ CO & Methanol & O$_2$ & 67.71 & 14.41 & 1.38 & 1.38 & 35.36 & 32.38 & 9.21 \\ CO & Methanol & N$_2$ & 74.72 & 13.80 & 1.40 & 1.40 & 38.24 & 35.37 & 8.11 \\ CO & Methanol & H$_2$S & 129.60 & 10.96 & 1.47 & 1.47 & 59.63 & 57.29 & 4.09 \\ CO & Methanol & NO & 72.38 & 13.87 & 1.40 & 1.40 & 37.13 & 34.25 & 8.42 \\ N$_2$O & Methanol & CH$_4$ & 147.94 & 12.34 & 1.43 & 1.44 & 71.84 & 67.87 & 5.85 \\ N$_2$O & Methanol & NO$_2$ & 171.27 & 13.90 & 1.40 & 1.40 & 87.95 & 80.97 & 8.62 \\ N$_2$O & Methanol & CO$_2$ & 174.42 & 13.92 & 1.40 & 1.40 & 89.62 & 82.64 & 8.45 \\ N$_2$O & Methanol & CO & 122.44 & 13.35 & 1.41 & 1.41 & 61.69 & 57.26 & 7.75 \\ N$_2$O & Methanol & N$_2$O & 188.52 & 13.58 & 1.40 & 1.40 & 95.73 & 88.60 & 8.05 \\ N$_2$O & Methanol & O$_3$ & 178.30 & 14.00 & 1.39 & 1.39 & 91.88 & 84.39 & 8.87 \\ N$_2$O & Methanol & O$_2$ & 104.35 & 14.65 & 1.38 & 1.38 & 54.94 & 50.17 & 9.51 \\ N$_2$O & Methanol & N$_2$ & 115.11 & 14.03 & 1.39 & 1.39 & 59.37 & 54.77 & 8.40 \\ N$_2$O & Methanol & H$_2$S & 199.10 & 11.13 & 1.47 & 1.47 & 92.22 & 88.38 & 4.34 \\ N$_2$O & Methanol & NO & 111.50 & 14.10 & 1.39 & 1.39 & 57.65 & 53.03 & 8.72 \\ O$_3$ & Methanol & CH$_4$ & 139.65 & 12.67 & 1.43 & 1.43 & 68.64 & 64.44 & 6.52 \\ O$_3$ & Methanol & NO$_2$ & 162.11 & 14.35 & 1.38 & 1.38 & 84.52 & 77.20 & 9.47 \\ O$_3$ & Methanol & CO$_2$ & 165.01 & 14.36 & 1.38 & 1.38 & 86.06 & 78.75 & 9.28 \\ O$_3$ & Methanol & CO & 115.77 & 13.76 & 1.40 & 1.40 & 59.17 & 54.51 & 8.55 \\ O$_3$ & Methanol & N$_2$O & 178.30 & 14.00 & 1.39 & 1.39 & 91.88 & 84.39 & 8.87 \\ O$_3$ & Methanol & O$_3$ & 168.83 & 14.47 & 1.38 & 1.38 & 88.35 & 80.51 & 9.73 \\ O$_3$ & Methanol & O$_2$ & 98.84 & 15.16 & 1.37 & 1.37 & 52.87 & 47.90 & 10.37 \\ O$_3$ & Methanol & N$_2$ & 108.90 & 14.48 & 1.38 & 1.38 & 57.01 & 52.20 & 9.21 \\ O$_3$ & Methanol & H$_2$S & 187.75 & 11.39 & 1.46 & 1.46 & 87.87 & 83.73 & 4.94 \\ O$_3$ & Methanol & NO & 105.53 & 14.56 & 1.38 & 1.38 & 55.40 & 50.57 & 9.56 \\ O$_2$ & Methanol & CH$_4$ & 81.51 & 13.24 & 1.41 & 1.41 & 40.91 & 38.18 & 7.16 \\ O$_2$ & Methanol & NO$_2$ & 94.92 & 15.03 & 1.37 & 1.37 & 50.57 & 45.93 & 10.11 \\ O$_2$ & Methanol & CO$_2$ & 96.69 & 15.03 & 1.37 & 1.37 & 51.51 & 46.88 & 9.89 \\ O$_2$ & Methanol & CO & 67.71 & 14.41 & 1.38 & 1.38 & 35.36 & 32.38 & 9.21 \\ O$_2$ & Methanol & N$_2$O & 104.35 & 14.65 & 1.38 & 1.38 & 54.94 & 50.17 & 9.51 \\ O$_2$ & Methanol & O$_3$ & 98.84 & 15.16 & 1.37 & 1.37 & 52.87 & 47.90 & 10.37 \\ O$_2$ & Methanol & O$_2$ & 58.01 & 15.86 & 1.35 & 1.35 & 31.70 & 28.57 & 10.94 \\ O$_2$ & Methanol & N$_2$ & 63.86 & 15.13 & 1.37 & 1.37 & 34.13 & 31.09 & 9.80 \\ O$_2$ & Methanol & H$_2$S & 109.06 & 11.92 & 1.45 & 1.45 & 52.12 & 49.35 & 5.61 \\ O$_2$ & Methanol & NO & 61.86 & 15.24 & 1.36 & 1.37 & 33.17 & 30.11 & 10.17 \\ N$_2$ & Methanol & CH$_4$ & 90.15 & 12.76 & 1.42 & 1.42 & 44.45 & 41.85 & 6.22 \\ N$_2$ & Methanol & NO$_2$ & 104.61 & 14.37 & 1.38 & 1.38 & 54.57 & 50.07 & 8.97 \\ N$_2$ & Methanol & CO$_2$ & 106.60 & 14.38 & 1.38 & 1.38 & 55.62 & 51.13 & 8.78 \\ N$_2$ & Methanol & CO & 74.72 & 13.80 & 1.40 & 1.40 & 38.24 & 35.37 & 8.11 \\ N$_2$ & Methanol & N$_2$O & 115.11 & 14.03 & 1.39 & 1.39 & 59.37 & 54.77 & 8.40 \\ N$_2$ & Methanol & O$_3$ & 108.90 & 14.48 & 1.38 & 1.38 & 57.01 & 52.20 & 9.21 \\ N$_2$ & Methanol & O$_2$ & 63.86 & 15.13 & 1.37 & 1.37 & 34.13 & 31.09 & 9.80 \\ N$_2$ & Methanol & N$_2$ & 70.40 & 14.49 & 1.38 & 1.38 & 36.86 & 33.91 & 8.72 \\ N$_2$ & Methanol & H$_2$S & 120.86 & 11.52 & 1.46 & 1.46 & 56.85 & 54.27 & 4.75 \\ N$_2$ & Methanol & NO & 68.17 & 14.57 & 1.38 & 1.38 & 35.79 & 32.82 & 9.05 \\ H$_2$S & Methanol & CH$_4$ & 157.89 & 10.28 & 1.49 & 1.50 & 70.63 & 68.76 & 2.71 \\ H$_2$S & Methanol & NO$_2$ & 180.42 & 11.34 & 1.46 & 1.46 & 84.27 & 80.42 & 4.78 \\ H$_2$S & Methanol & CO$_2$ & 183.51 & 11.40 & 1.46 & 1.46 & 85.93 & 82.06 & 4.71 \\ H$_2$S & Methanol & CO & 129.60 & 10.96 & 1.47 & 1.47 & 59.63 & 57.29 & 4.09 \\ H$_2$S & Methanol & N$_2$O & 199.10 & 11.13 & 1.47 & 1.47 & 92.22 & 88.38 & 4.34 \\ H$_2$S & Methanol & O$_3$ & 187.75 & 11.39 & 1.46 & 1.46 & 87.87 & 83.73 & 4.94 \\ H$_2$S & Methanol & O$_2$ & 109.06 & 11.92 & 1.45 & 1.45 & 52.12 & 49.35 & 5.61 \\ H$_2$S & Methanol & N$_2$ & 120.86 & 11.52 & 1.46 & 1.46 & 56.85 & 54.27 & 4.75 \\ H$_2$S & Methanol & H$_2$S & 215.59 & 9.33 & 1.52 & 1.53 & 92.60 & 91.23 & 1.50 \\ H$_2$S & Methanol & NO & 117.12 & 11.53 & 1.46 & 1.46 & 55.11 & 52.52 & 4.94 \\ NO & Methanol & CH$_4$ & 87.30 & 12.79 & 1.42 & 1.42 & 43.10 & 40.49 & 6.46 \\ NO & Methanol & NO$_2$ & 101.36 & 14.45 & 1.38 & 1.38 & 53.02 & 48.50 & 9.31 \\ NO & Methanol & CO$_2$ & 103.25 & 14.46 & 1.38 & 1.38 & 54.02 & 49.50 & 9.11 \\ NO & Methanol & CO & 72.38 & 13.87 & 1.40 & 1.40 & 37.13 & 34.25 & 8.42 \\ NO & Methanol & N$_2$O & 111.50 & 14.10 & 1.39 & 1.39 & 57.65 & 53.03 & 8.72 \\ NO & Methanol & O$_3$ & 105.53 & 14.56 & 1.38 & 1.38 & 55.40 & 50.57 & 9.56 \\ NO & Methanol & O$_2$ & 61.86 & 15.24 & 1.36 & 1.37 & 33.17 & 30.11 & 10.17 \\ NO & Methanol & N$_2$ & 68.17 & 14.57 & 1.38 & 1.38 & 35.79 & 32.82 & 9.05 \\ NO & Methanol & H$_2$S & 117.12 & 11.53 & 1.46 & 1.46 & 55.11 & 52.52 & 4.94 \\ NO & Methanol & NO & 66.03 & 14.66 & 1.38 & 1.38 & 34.77 & 31.78 & 9.39 \\ 

CH$_4$ & Butanol & CH$_4$ & 116.68 & 11.35 & 1.56 & 1.56 & 48.06 & 46.41 & 3.56 \\ CH$_4$ & Butanol & NO$_2$ & 134.21 & 12.60 & 1.52 & 1.52 & 58.34 & 55.10 & 5.89 \\ CH$_4$ & Butanol & CO$_2$ & 136.68 & 12.64 & 1.52 & 1.52 & 59.53 & 56.27 & 5.78 \\ CH$_4$ & Butanol & CO & 96.18 & 12.15 & 1.53 & 1.53 & 41.04 & 39.03 & 5.15 \\ CH$_4$ & Butanol & N$_2$O & 147.94 & 12.34 & 1.53 & 1.52 & 63.63 & 60.36 & 5.41 \\ CH$_4$ & Butanol & O$_3$ & 139.65 & 12.67 & 1.52 & 1.51 & 60.89 & 57.40 & 6.08 \\ CH$_4$ & Butanol & O$_2$ & 81.51 & 13.24 & 1.50 & 1.50 & 36.38 & 34.09 & 6.71 \\ CH$_4$ & Butanol & N$_2$ & 90.15 & 12.76 & 1.51 & 1.51 & 39.45 & 37.30 & 5.77 \\ CH$_4$ & Butanol & H$_2$S & 157.89 & 10.28 & 1.60 & 1.60 & 61.87 & 60.55 & 2.18 \\ CH$_4$ & Butanol & NO & 87.30 & 12.79 & 1.51 & 1.51 & 38.26 & 36.09 & 6.02 \\ NO$_2$ & Butanol & CH$_4$ & 134.21 & 12.60 & 1.52 & 1.52 & 58.34 & 55.10 & 5.89 \\ NO$_2$ & Butanol & NO$_2$ & 155.68 & 14.24 & 1.47 & 1.47 & 72.17 & 66.36 & 8.76 \\ NO$_2$ & Butanol & CO$_2$ & 158.50 & 14.26 & 1.47 & 1.47 & 73.51 & 67.71 & 8.57 \\ NO$_2$ & Butanol & CO & 111.21 & 13.67 & 1.48 & 1.48 & 50.46 & 46.77 & 7.89 \\ NO$_2$ & Butanol & N$_2$O & 171.27 & 13.90 & 1.48 & 1.48 & 78.40 & 72.46 & 8.18 \\ NO$_2$ & Butanol & O$_3$ & 162.11 & 14.35 & 1.46 & 1.47 & 75.45 & 69.22 & 9.00 \\ NO$_2$ & Butanol & O$_2$ & 94.92 & 15.03 & 1.44 & 1.45 & 45.24 & 41.29 & 9.58 \\ NO$_2$ & Butanol & N$_2$ & 104.61 & 14.37 & 1.46 & 1.47 & 48.72 & 44.91 & 8.48 \\ NO$_2$ & Butanol & H$_2$S & 180.42 & 11.34 & 1.56 & 1.56 & 74.27 & 71.18 & 4.35 \\ NO$_2$ & Butanol & NO & 101.36 & 14.45 & 1.46 & 1.46 & 47.35 & 43.51 & 8.83 \\ CO$_2$ & Butanol & CH$_4$ & 136.68 & 12.64 & 1.52 & 1.52 & 59.53 & 56.27 & 5.78 \\ CO$_2$ & Butanol & NO$_2$ & 158.50 & 14.26 & 1.47 & 1.47 & 73.51 & 67.71 & 8.57 \\ CO$_2$ & Butanol & CO$_2$ & 161.47 & 14.27 & 1.47 & 1.47 & 74.92 & 69.13 & 8.38 \\ CO$_2$ & Butanol & CO & 113.25 & 13.69 & 1.48 & 1.48 & 51.42 & 47.74 & 7.73 \\ CO$_2$ & Butanol & N$_2$O & 174.42 & 13.92 & 1.48 & 1.48 & 79.89 & 73.97 & 8.01 \\ CO$_2$ & Butanol & O$_3$ & 165.01 & 14.36 & 1.46 & 1.47 & 76.83 & 70.61 & 8.80 \\ CO$_2$ & Butanol & O$_2$ & 96.69 & 15.03 & 1.44 & 1.45 & 46.08 & 42.14 & 9.36 \\ CO$_2$ & Butanol & N$_2$ & 106.60 & 14.38 & 1.46 & 1.47 & 49.66 & 45.86 & 8.29 \\ CO$_2$ & Butanol & H$_2$S & 183.51 & 11.40 & 1.56 & 1.56 & 75.76 & 72.64 & 4.29 \\ CO$_2$ & Butanol & NO & 103.25 & 14.46 & 1.46 & 1.46 & 48.24 & 44.41 & 8.63 \\ CO & Butanol & CH$_4$ & 96.18 & 12.15 & 1.53 & 1.53 & 41.04 & 39.03 & 5.15 \\ CO & Butanol & NO$_2$ & 111.21 & 13.67 & 1.48 & 1.48 & 50.46 & 46.77 & 7.89 \\ CO & Butanol & CO$_2$ & 113.25 & 13.69 & 1.48 & 1.48 & 51.42 & 47.74 & 7.73 \\ CO & Butanol & CO & 79.54 & 13.13 & 1.50 & 1.50 & 35.35 & 33.02 & 7.05 \\ CO & Butanol & N$_2$O & 122.44 & 13.35 & 1.49 & 1.49 & 54.88 & 51.13 & 7.34 \\ CO & Butanol & O$_3$ & 115.77 & 13.76 & 1.48 & 1.48 & 52.72 & 48.76 & 8.13 \\ CO & Butanol & O$_2$ & 67.71 & 14.41 & 1.46 & 1.46 & 31.58 & 29.04 & 8.73 \\ CO & Butanol & N$_2$ & 74.72 & 13.80 & 1.48 & 1.48 & 34.08 & 31.65 & 7.66 \\ CO & Butanol & H$_2$S & 129.60 & 10.96 & 1.57 & 1.57 & 52.45 & 50.60 & 3.64 \\ CO & Butanol & NO & 72.38 & 13.87 & 1.48 & 1.48 & 33.10 & 30.65 & 7.98 \\ N$_2$O & Butanol & CH$_4$ & 147.94 & 12.34 & 1.53 & 1.52 & 63.63 & 60.36 & 5.41 \\ N$_2$O & Butanol & NO$_2$ & 171.27 & 13.90 & 1.48 & 1.48 & 78.40 & 72.46 & 8.18 \\ N$_2$O & Butanol & CO$_2$ & 174.42 & 13.92 & 1.48 & 1.48 & 79.89 & 73.97 & 8.01 \\ N$_2$O & Butanol & CO & 122.44 & 13.35 & 1.49 & 1.49 & 54.88 & 51.13 & 7.34 \\ N$_2$O & Butanol & N$_2$O & 188.52 & 13.58 & 1.49 & 1.49 & 85.23 & 79.19 & 7.63 \\ N$_2$O & Butanol & O$_3$ & 178.30 & 14.00 & 1.47 & 1.48 & 81.92 & 75.56 & 8.42 \\ N$_2$O & Butanol & O$_2$ & 104.35 & 14.65 & 1.45 & 1.46 & 49.09 & 45.04 & 9.01 \\ N$_2$O & Butanol & N$_2$ & 115.11 & 14.03 & 1.47 & 1.47 & 52.95 & 49.05 & 7.93 \\ N$_2$O & Butanol & H$_2$S & 199.10 & 11.13 & 1.57 & 1.57 & 81.18 & 78.13 & 3.90 \\ N$_2$O & Butanol & NO & 111.50 & 14.10 & 1.47 & 1.47 & 51.43 & 47.50 & 8.26 \\ O$_3$ & Butanol & CH$_4$ & 139.65 & 12.67 & 1.52 & 1.51 & 60.89 & 57.40 & 6.08 \\ O$_3$ & Butanol & NO$_2$ & 162.11 & 14.35 & 1.46 & 1.47 & 75.45 & 69.22 & 9.00 \\ O$_3$ & Butanol & CO$_2$ & 165.01 & 14.36 & 1.46 & 1.47 & 76.83 & 70.61 & 8.80 \\ O$_3$ & Butanol & CO & 115.77 & 13.76 & 1.48 & 1.48 & 52.72 & 48.76 & 8.13 \\ O$_3$ & Butanol & N$_2$O & 178.30 & 14.00 & 1.47 & 1.48 & 81.92 & 75.56 & 8.42 \\ O$_3$ & Butanol & O$_3$ & 168.83 & 14.47 & 1.46 & 1.46 & 78.90 & 72.22 & 9.25 \\ O$_3$ & Butanol & O$_2$ & 98.84 & 15.16 & 1.44 & 1.45 & 47.32 & 43.08 & 9.82 \\ O$_3$ & Butanol & N$_2$ & 108.90 & 14.48 & 1.46 & 1.46 & 50.91 & 46.83 & 8.71 \\ O$_3$ & Butanol & H$_2$S & 187.75 & 11.39 & 1.56 & 1.56 & 77.46 & 74.12 & 4.51 \\ O$_3$ & Butanol & NO & 105.53 & 14.56 & 1.46 & 1.46 & 49.49 & 45.38 & 9.07 \\ O$_2$ & Butanol & CH$_4$ & 81.51 & 13.24 & 1.50 & 1.50 & 36.38 & 34.09 & 6.71 \\ O$_2$ & Butanol & NO$_2$ & 94.92 & 15.03 & 1.44 & 1.45 & 45.24 & 41.29 & 9.58 \\ O$_2$ & Butanol & CO$_2$ & 96.69 & 15.03 & 1.44 & 1.45 & 46.08 & 42.14 & 9.36 \\ O$_2$ & Butanol & CO & 67.71 & 14.41 & 1.46 & 1.46 & 31.58 & 29.04 & 8.73 \\ O$_2$ & Butanol & N$_2$O & 104.35 & 14.65 & 1.45 & 1.46 & 49.09 & 45.04 & 9.01 \\ O$_2$ & Butanol & O$_3$ & 98.84 & 15.16 & 1.44 & 1.45 & 47.32 & 43.08 & 9.82 \\ O$_2$ & Butanol & O$_2$ & 58.01 & 15.86 & 1.42 & 1.43 & 28.42 & 25.76 & 10.31 \\ O$_2$ & Butanol & N$_2$ & 63.86 & 15.13 & 1.44 & 1.45 & 30.55 & 27.96 & 9.24 \\ O$_2$ & Butanol & H$_2$S & 109.06 & 11.92 & 1.54 & 1.54 & 46.08 & 43.79 & 5.21 \\ O$_2$ & Butanol & NO & 61.86 & 15.24 & 1.44 & 1.44 & 29.69 & 27.09 & 9.61 \\ N$_2$ & Butanol & CH$_4$ & 90.15 & 12.76 & 1.51 & 1.51 & 39.45 & 37.30 & 5.77 \\ N$_2$ & Butanol & NO$_2$ & 104.61 & 14.37 & 1.46 & 1.47 & 48.72 & 44.91 & 8.48 \\ N$_2$ & Butanol & CO$_2$ & 106.60 & 14.38 & 1.46 & 1.47 & 49.66 & 45.86 & 8.29 \\ N$_2$ & Butanol & CO & 74.72 & 13.80 & 1.48 & 1.48 & 34.08 & 31.65 & 7.66 \\ N$_2$ & Butanol & N$_2$O & 115.11 & 14.03 & 1.47 & 1.47 & 52.95 & 49.05 & 7.93 \\ N$_2$ & Butanol & O$_3$ & 108.90 & 14.48 & 1.46 & 1.46 & 50.91 & 46.83 & 8.71 \\ N$_2$ & Butanol & O$_2$ & 63.86 & 15.13 & 1.44 & 1.45 & 30.55 & 27.96 & 9.24 \\ N$_2$ & Butanol & N$_2$ & 70.40 & 14.49 & 1.46 & 1.46 & 32.92 & 30.43 & 8.20 \\ N$_2$ & Butanol & H$_2$S & 120.86 & 11.52 & 1.55 & 1.55 & 50.15 & 48.08 & 4.32 \\ N$_2$ & Butanol & NO & 68.17 & 14.57 & 1.46 & 1.46 & 31.97 & 29.46 & 8.53 \\ H$_2$S & Butanol & CH$_4$ & 157.89 & 10.28 & 1.60 & 1.60 & 61.87 & 60.55 & 2.18 \\ H$_2$S & Butanol & NO$_2$ & 180.42 & 11.34 & 1.56 & 1.56 & 74.27 & 71.18 & 4.35 \\ H$_2$S & Butanol & CO$_2$ & 183.51 & 11.40 & 1.56 & 1.56 & 75.76 & 72.64 & 4.29 \\ H$_2$S & Butanol & CO & 129.60 & 10.96 & 1.57 & 1.57 & 52.45 & 50.60 & 3.64 \\ H$_2$S & Butanol & N$_2$O & 199.10 & 11.13 & 1.57 & 1.57 & 81.18 & 78.13 & 3.90 \\ H$_2$S & Butanol & O$_3$ & 187.75 & 11.39 & 1.56 & 1.56 & 77.46 & 74.12 & 4.51 \\ H$_2$S & Butanol & O$_2$ & 109.06 & 11.92 & 1.54 & 1.54 & 46.08 & 43.79 & 5.21 \\ H$_2$S & Butanol & N$_2$ & 120.86 & 11.52 & 1.55 & 1.55 & 50.15 & 48.08 & 4.32 \\ H$_2$S & Butanol & H$_2$S & 215.59 & 9.33 & 1.64 & 1.64 & 80.60 & 79.90 & 0.87 \\ H$_2$S & Butanol & NO & 117.12 & 11.53 & 1.55 & 1.55 & 48.62 & 46.52 & 4.52 \\ NO & Butanol & CH$_4$ & 87.30 & 12.79 & 1.51 & 1.51 & 38.26 & 36.09 & 6.02 \\ NO & Butanol & NO$_2$ & 101.36 & 14.45 & 1.46 & 1.46 & 47.35 & 43.51 & 8.83 \\ NO & Butanol & CO$_2$ & 103.25 & 14.46 & 1.46 & 1.46 & 48.24 & 44.41 & 8.63 \\ NO & Butanol & CO & 72.38 & 13.87 & 1.48 & 1.48 & 33.10 & 30.65 & 7.98 \\ NO & Butanol & N$_2$O & 111.50 & 14.10 & 1.47 & 1.47 & 51.43 & 47.50 & 8.26 \\ NO & Butanol & O$_3$ & 105.53 & 14.56 & 1.46 & 1.46 & 49.49 & 45.38 & 9.07 \\ NO & Butanol & O$_2$ & 61.86 & 15.24 & 1.44 & 1.44 & 29.69 & 27.09 & 9.61 \\ NO & Butanol & N$_2$ & 68.17 & 14.57 & 1.46 & 1.46 & 31.97 & 29.46 & 8.53 \\ NO & Butanol & H$_2$S & 117.12 & 11.53 & 1.55 & 1.55 & 48.62 & 46.52 & 4.52 \\ NO & Butanol & NO & 66.03 & 14.66 & 1.45 & 1.46 & 31.07 & 28.53 & 8.88 \\ 

CH$_4$ & Propanol & CH$_4$ & 116.68 & 11.35 & 1.54 & 1.54 & 49.31 & 47.35 & 4.14 \\ CH$_4$ & Propanol & NO$_2$ & 134.21 & 12.60 & 1.50 & 1.50 & 59.81 & 56.16 & 6.50 \\ CH$_4$ & Propanol & CO$_2$ & 136.68 & 12.64 & 1.50 & 1.50 & 61.02 & 57.36 & 6.38 \\ CH$_4$ & Propanol & CO & 96.18 & 12.15 & 1.51 & 1.51 & 42.08 & 39.80 & 5.74 \\ CH$_4$ & Propanol & N$_2$O & 147.94 & 12.34 & 1.51 & 1.51 & 65.25 & 61.54 & 6.01 \\ CH$_4$ & Propanol & O$_3$ & 139.65 & 12.67 & 1.50 & 1.50 & 62.42 & 58.50 & 6.70 \\ CH$_4$ & Propanol & O$_2$ & 81.51 & 13.24 & 1.48 & 1.48 & 37.28 & 34.74 & 7.32 \\ CH$_4$ & Propanol & N$_2$ & 90.15 & 12.76 & 1.49 & 1.49 & 40.44 & 38.02 & 6.37 \\ CH$_4$ & Propanol & H$_2$S & 157.89 & 10.28 & 1.58 & 1.58 & 63.53 & 61.82 & 2.76 \\ CH$_4$ & Propanol & NO & 87.30 & 12.79 & 1.49 & 1.49 & 39.22 & 36.78 & 6.62 \\ NO$_2$ & Propanol & CH$_4$ & 134.21 & 12.60 & 1.50 & 1.50 & 59.81 & 56.16 & 6.50 \\ NO$_2$ & Propanol & NO$_2$ & 155.68 & 14.24 & 1.45 & 1.45 & 73.92 & 67.57 & 9.40 \\ NO$_2$ & Propanol & CO$_2$ & 158.50 & 14.26 & 1.45 & 1.45 & 75.29 & 68.95 & 9.19 \\ NO$_2$ & Propanol & CO & 111.21 & 13.67 & 1.47 & 1.47 & 51.69 & 47.64 & 8.52 \\ NO$_2$ & Propanol & N$_2$O & 171.27 & 13.90 & 1.46 & 1.46 & 80.31 & 73.80 & 8.81 \\ NO$_2$ & Propanol & O$_3$ & 162.11 & 14.35 & 1.45 & 1.45 & 77.27 & 70.47 & 9.65 \\ NO$_2$ & Propanol & O$_2$ & 94.92 & 15.03 & 1.43 & 1.43 & 46.32 & 42.02 & 10.22 \\ NO$_2$ & Propanol & N$_2$ & 104.61 & 14.37 & 1.45 & 1.45 & 49.89 & 45.73 & 9.10 \\ NO$_2$ & Propanol & H$_2$S & 180.42 & 11.34 & 1.54 & 1.54 & 76.21 & 72.61 & 4.96 \\ NO$_2$ & Propanol & NO & 101.36 & 14.45 & 1.44 & 1.45 & 48.48 & 44.29 & 9.46 \\ CO$_2$ & Propanol & CH$_4$ & 136.68 & 12.64 & 1.50 & 1.50 & 61.02 & 57.36 & 6.38 \\ CO$_2$ & Propanol & NO$_2$ & 158.50 & 14.26 & 1.45 & 1.45 & 75.29 & 68.95 & 9.19 \\ CO$_2$ & Propanol & CO$_2$ & 161.47 & 14.27 & 1.45 & 1.45 & 76.73 & 70.40 & 8.99 \\ CO$_2$ & Propanol & CO & 113.25 & 13.69 & 1.47 & 1.47 & 52.68 & 48.63 & 8.34 \\ CO$_2$ & Propanol & N$_2$O & 174.42 & 13.92 & 1.46 & 1.46 & 81.84 & 75.34 & 8.62 \\ CO$_2$ & Propanol & O$_3$ & 165.01 & 14.36 & 1.45 & 1.45 & 78.68 & 71.90 & 9.44 \\ CO$_2$ & Propanol & O$_2$ & 96.69 & 15.03 & 1.43 & 1.43 & 47.18 & 42.89 & 9.99 \\ CO$_2$ & Propanol & N$_2$ & 106.60 & 14.38 & 1.45 & 1.45 & 50.86 & 46.70 & 8.90 \\ CO$_2$ & Propanol & H$_2$S & 183.51 & 11.40 & 1.54 & 1.54 & 77.73 & 74.11 & 4.88 \\ CO$_2$ & Propanol & NO & 103.25 & 14.46 & 1.44 & 1.45 & 49.40 & 45.22 & 9.25 \\ CO & Propanol & CH$_4$ & 96.18 & 12.15 & 1.51 & 1.51 & 42.08 & 39.80 & 5.74 \\ CO & Propanol & NO$_2$ & 111.21 & 13.67 & 1.47 & 1.47 & 51.69 & 47.64 & 8.52 \\ CO & Propanol & CO$_2$ & 113.25 & 13.69 & 1.47 & 1.47 & 52.68 & 48.63 & 8.34 \\ CO & Propanol & CO & 79.54 & 13.13 & 1.48 & 1.48 & 36.22 & 33.65 & 7.66 \\ CO & Propanol & N$_2$O & 122.44 & 13.35 & 1.48 & 1.48 & 56.24 & 52.09 & 7.95 \\ CO & Propanol & O$_3$ & 115.77 & 13.76 & 1.46 & 1.46 & 54.01 & 49.66 & 8.76 \\ CO & Propanol & O$_2$ & 67.71 & 14.41 & 1.44 & 1.45 & 32.34 & 29.57 & 9.36 \\ CO & Propanol & N$_2$ & 74.72 & 13.80 & 1.46 & 1.46 & 34.91 & 32.24 & 8.27 \\ CO & Propanol & H$_2$S & 129.60 & 10.96 & 1.55 & 1.55 & 53.83 & 51.64 & 4.23 \\ CO & Propanol & NO & 72.38 & 13.87 & 1.46 & 1.46 & 33.90 & 31.22 & 8.60 \\ N$_2$O & Propanol & CH$_4$ & 147.94 & 12.34 & 1.51 & 1.51 & 65.25 & 61.54 & 6.01 \\ N$_2$O & Propanol & NO$_2$ & 171.27 & 13.90 & 1.46 & 1.46 & 80.31 & 73.80 & 8.81 \\ N$_2$O & Propanol & CO$_2$ & 174.42 & 13.92 & 1.46 & 1.46 & 81.84 & 75.34 & 8.62 \\ N$_2$O & Propanol & CO & 122.44 & 13.35 & 1.48 & 1.48 & 56.24 & 52.09 & 7.95 \\ N$_2$O & Propanol & N$_2$O & 188.52 & 13.58 & 1.47 & 1.47 & 87.33 & 80.68 & 8.24 \\ N$_2$O & Propanol & O$_3$ & 178.30 & 14.00 & 1.46 & 1.46 & 83.92 & 76.95 & 9.05 \\ N$_2$O & Propanol & O$_2$ & 104.35 & 14.65 & 1.44 & 1.44 & 50.27 & 45.85 & 9.64 \\ N$_2$O & Propanol & N$_2$ & 115.11 & 14.03 & 1.46 & 1.46 & 54.23 & 49.96 & 8.55 \\ N$_2$O & Propanol & H$_2$S & 199.10 & 11.13 & 1.55 & 1.55 & 83.31 & 79.73 & 4.50 \\ N$_2$O & Propanol & NO & 111.50 & 14.10 & 1.45 & 1.45 & 52.67 & 48.38 & 8.89 \\ O$_3$ & Propanol & CH$_4$ & 139.65 & 12.67 & 1.50 & 1.50 & 62.42 & 58.50 & 6.70 \\ O$_3$ & Propanol & NO$_2$ & 162.11 & 14.35 & 1.45 & 1.45 & 77.27 & 70.47 & 9.65 \\ O$_3$ & Propanol & CO$_2$ & 165.01 & 14.36 & 1.45 & 1.45 & 78.68 & 71.90 & 9.44 \\ O$_3$ & Propanol & CO & 115.77 & 13.76 & 1.46 & 1.46 & 54.01 & 49.66 & 8.76 \\ O$_3$ & Propanol & N$_2$O & 178.30 & 14.00 & 1.46 & 1.46 & 83.92 & 76.95 & 9.05 \\ O$_3$ & Propanol & O$_3$ & 168.83 & 14.47 & 1.44 & 1.45 & 80.80 & 73.52 & 9.90 \\ O$_3$ & Propanol & O$_2$ & 98.84 & 15.16 & 1.42 & 1.43 & 48.44 & 43.84 & 10.47 \\ O$_3$ & Propanol & N$_2$ & 108.90 & 14.48 & 1.44 & 1.45 & 52.13 & 47.68 & 9.34 \\ O$_3$ & Propanol & H$_2$S & 187.75 & 11.39 & 1.54 & 1.54 & 79.48 & 75.61 & 5.12 \\ O$_3$ & Propanol & NO & 105.53 & 14.56 & 1.44 & 1.44 & 50.68 & 46.20 & 9.71 \\ O$_2$ & Propanol & CH$_4$ & 81.51 & 13.24 & 1.48 & 1.48 & 37.28 & 34.74 & 7.32 \\ O$_2$ & Propanol & NO$_2$ & 94.92 & 15.03 & 1.43 & 1.43 & 46.32 & 42.02 & 10.22 \\ O$_2$ & Propanol & CO$_2$ & 96.69 & 15.03 & 1.43 & 1.43 & 47.18 & 42.89 & 9.99 \\ O$_2$ & Propanol & CO & 67.71 & 14.41 & 1.44 & 1.45 & 32.34 & 29.57 & 9.36 \\ O$_2$ & Propanol & N$_2$O & 104.35 & 14.65 & 1.44 & 1.44 & 50.27 & 45.85 & 9.64 \\ O$_2$ & Propanol & O$_3$ & 98.84 & 15.16 & 1.42 & 1.43 & 48.44 & 43.84 & 10.47 \\ O$_2$ & Propanol & O$_2$ & 58.01 & 15.86 & 1.41 & 1.41 & 29.08 & 26.21 & 10.96 \\ O$_2$ & Propanol & N$_2$ & 63.86 & 15.13 & 1.43 & 1.43 & 31.27 & 28.46 & 9.87 \\ O$_2$ & Propanol & H$_2$S & 109.06 & 11.92 & 1.52 & 1.52 & 47.26 & 44.66 & 5.81 \\ O$_2$ & Propanol & NO & 61.86 & 15.24 & 1.42 & 1.43 & 30.40 & 27.57 & 10.24 \\ N$_2$ & Propanol & CH$_4$ & 90.15 & 12.76 & 1.49 & 1.49 & 40.44 & 38.02 & 6.37 \\ N$_2$ & Propanol & NO$_2$ & 104.61 & 14.37 & 1.45 & 1.45 & 49.89 & 45.73 & 9.10 \\ N$_2$ & Propanol & CO$_2$ & 106.60 & 14.38 & 1.45 & 1.45 & 50.86 & 46.70 & 8.90 \\ N$_2$ & Propanol & CO & 74.72 & 13.80 & 1.46 & 1.46 & 34.91 & 32.24 & 8.27 \\ N$_2$ & Propanol & N$_2$O & 115.11 & 14.03 & 1.46 & 1.46 & 54.23 & 49.96 & 8.55 \\ N$_2$ & Propanol & O$_3$ & 108.90 & 14.48 & 1.44 & 1.45 & 52.13 & 47.68 & 9.34 \\ N$_2$ & Propanol & O$_2$ & 63.86 & 15.13 & 1.43 & 1.43 & 31.27 & 28.46 & 9.87 \\ N$_2$ & Propanol & N$_2$ & 70.40 & 14.49 & 1.44 & 1.44 & 33.71 & 30.98 & 8.82 \\ N$_2$ & Propanol & H$_2$S & 120.86 & 11.52 & 1.53 & 1.53 & 51.45 & 49.05 & 4.91 \\ N$_2$ & Propanol & NO & 68.17 & 14.57 & 1.44 & 1.44 & 32.74 & 30.00 & 9.16 \\ H$_2$S & Propanol & CH$_4$ & 157.89 & 10.28 & 1.58 & 1.58 & 63.53 & 61.82 & 2.76 \\ H$_2$S & Propanol & NO$_2$ & 180.42 & 11.34 & 1.54 & 1.54 & 76.21 & 72.61 & 4.96 \\ H$_2$S & Propanol & CO$_2$ & 183.51 & 11.40 & 1.54 & 1.54 & 77.73 & 74.11 & 4.88 \\ H$_2$S & Propanol & CO & 129.60 & 10.96 & 1.55 & 1.55 & 53.83 & 51.64 & 4.23 \\ H$_2$S & Propanol & N$_2$O & 199.10 & 11.13 & 1.55 & 1.55 & 83.31 & 79.73 & 4.50 \\ H$_2$S & Propanol & O$_3$ & 187.75 & 11.39 & 1.54 & 1.54 & 79.48 & 75.61 & 5.12 \\ H$_2$S & Propanol & O$_2$ & 109.06 & 11.92 & 1.52 & 1.52 & 47.26 & 44.66 & 5.81 \\ H$_2$S & Propanol & N$_2$ & 120.86 & 11.52 & 1.53 & 1.53 & 51.45 & 49.05 & 4.91 \\ H$_2$S & Propanol & H$_2$S & 215.59 & 9.33 & 1.61 & 1.61 & 82.84 & 81.64 & 1.47 \\ H$_2$S & Propanol & NO & 117.12 & 11.53 & 1.53 & 1.53 & 49.89 & 47.46 & 5.12 \\ NO & Propanol & CH$_4$ & 87.30 & 12.79 & 1.49 & 1.49 & 39.22 & 36.78 & 6.62 \\ NO & Propanol & NO$_2$ & 101.36 & 14.45 & 1.44 & 1.45 & 48.48 & 44.29 & 9.46 \\ NO & Propanol & CO$_2$ & 103.25 & 14.46 & 1.44 & 1.45 & 49.40 & 45.22 & 9.25 \\ NO & Propanol & CO & 72.38 & 13.87 & 1.46 & 1.46 & 33.90 & 31.22 & 8.60 \\ NO & Propanol & N$_2$O & 111.50 & 14.10 & 1.45 & 1.45 & 52.67 & 48.38 & 8.89 \\ NO & Propanol & O$_3$ & 105.53 & 14.56 & 1.44 & 1.44 & 50.68 & 46.20 & 9.71 \\ NO & Propanol & O$_2$ & 61.86 & 15.24 & 1.42 & 1.43 & 30.40 & 27.57 & 10.24 \\ NO & Propanol & N$_2$ & 68.17 & 14.57 & 1.44 & 1.44 & 32.74 & 30.00 & 9.16 \\ NO & Propanol & H$_2$S & 117.12 & 11.53 & 1.53 & 1.53 & 49.89 & 47.46 & 5.12 \\ NO & Propanol & NO & 66.03 & 14.66 & 1.44 & 1.44 & 31.81 & 29.05 & 9.51 \\

\end{longtable}
\end{document}